\documentclass{book} \usepackage{amsmath} \usepackage{hhline}
\usepackage{graphicx} \usepackage[T2A]{fontenc}
\usepackage[cp1251]{inputenc} \usepackage[russian]{babel}

\begin{document}

\def\ds{\displaystyle} \def\ss{\scriptsize} \def\hh{\hskip 1pt}
\def\hs{\hskip 2pt} \def\h{\hskip 0.2mm} \def\pr{\prime}

\newcommand{\fbR}{\hbox{\ss\bf R}}
\newcommand{\fbr}{\hbox{\footnotesize\bf r}}
\newcommand{\fbk}{\hbox{\footnotesize\bf k}}
\newcommand{\Tr}{\mathop{\rm T\h r}\nolimits}

\parindent=5mm

\setcounter{page}{1}

\makeatletter\renewcommand{\@evenhead} {\vbox{\hbox
to\textwidth{\thepage\hfil\small\it Б.В.Бондарев \hfil}\vskip
1mm\hrule\vskip -6mm}}

\renewcommand{\@evenhead}{}

\makeatletter\renewcommand{\@oddhead} {\vbox{\hbox
to\textwidth{\hfill\small\it Функция Ферми-Дирика и энергетическая щель \hfill\rm\thepage}\vskip 1mm\hrule\vskip -6mm}}

\renewcommand{\@oddhead}{}

\phantom{X} \vskip 25mm

\centerline {\bf\Large NEW THEORY OF SUPERCONDUCTIVITY. } \vskip
1mm \centerline {\bf\Large METHOD OF EQUILIBRIUM DENSITY MATRIX }
\vskip 7mm

\centerline {\large\bf Boris V. BONDAREV } \vskip 7mm

\centerline{\it Moscow Aviation Institute, Volokolamskoye Shosse 4,
125871, Moscow, Russia }\vskip 7mm

\centerline{E\hh -mail: bondarev.b@mail.ru }\vskip 7mm

\par{\it A new variational method for studying the equilibrium states of an interacting particles system has been proposed. The statistical description of the system is realized by means of a density matrix. This method is used for description of conduction electrons in metals. An integral equation for the electron distribution function over wave vectors has been obtained. The solutions of this equation have been found for those cases where the single-particle Hamiltonian and the electron interaction Hamiltonian can be approximated by a quite simple expression. It is shown that the distribution function at temperatures below the critical value possesses previously unknown features which allow to explain the superconductivity of metals and presence of a gap in the energy spectrum of superconducting electrons. } \vskip 7mm

\centerline {\bf Introduction}\vskip 2mm

\par Cooperative phenomena, such as superconductivity, ferro- and antiferromagnetism, etc., can be correctly explained quantitatively only in the framework of the quantum theory of many-particle systems. The most general statistical description of the studied system is carried out in quantum mechanics by means of the density matrix [1-4]. Such description is valid for equilibrium as well as for non-equilibrium systems, for both closed dynamical systems and systems that interact with their environment. If the studied system is in contact with a heat reservoir, then, strictly speaking, its description by means of the density matrix will be exclusively correct.

\par A complete statistical description of the system consisting of $N$ identical particles (say, fermions) in quantum mechanics is given by the statistical operator $\hat\varrho^{(N)}$ which satisfies the normalization condition
$$ \Tr_{12...N}\hh\hat\varrho^{(N)}=N!\hh . $$
By means of this operator, it is possible to build a hierarchical sequence of operators $\hat\varrho^{(1)}$, $\hat\varrho^{(2)}$, ... , defined by the relation
$$ \hat\varrho^{(n)}=\frac{1}{(N-n)!}\hs
\Tr_{n+1...N}\hh\hat\varrho^{(N)}\hh , $$
where $n=1$, 2,... , $N-1$. Although the statistical operators of lower orders give not a complete but reduced description of the many-particle systems, they are irreplaceable due to their simplicity in those cases where it is necessary to obtain practically useful equations and expressions. The implementation of the reduced description is supported by the fact, that all the observed physical quantities characterizing the state of a macroscopic system can be expressed exactly or approximately by the operators $\hat\varrho^{(1)}$ and $\hat\varrho^{(2)}$ or only by the single-particle operator $\hat\varrho^{(1)}$.

\par For a system in statistical equilibrium, the $N$-particle statistical operator has the form:
$$ \hat\varrho^{(N)}=\frac{1}{Z}\hs\exp\Bigl(-\hs\beta\hs\hat H^{(N)}\Bigr)\hh ,
\eqno (1) $$
where $\hat H^{(N)}$ is the Hamiltonian of the system, $\beta$ is the inverse temperature: $\beta=(k_B\hh T)^{\h -\h 1}$; $Z$ is the partition function:
$$ Z=\frac{1}{N!}\hs\Tr_{12...N}\hs\exp\Bigl(-\hs\beta\hs\hat H^{(N)}\Bigr)\hh . $$
Equation (1) is the only case where the many-particle statistical operator can be known.
\par The thermodynamic functions of state for the equilibrium system can be determined by means of the Gibbs method using the partition function. However, calculation of the partition function for the system of interacting particles is a very laborious problem that can be solved exactly only in some rare cases. Even approximate calculations of the partition function are very difficult.

\par The single-particle statistical operator can be found by the formula
$$ \hat\varrho^{(1)}=\frac{1}{(N-1)!}\hs
\Tr_{2...N}\hh\hat\varrho^{(N)}\hh . $$
But the operation $\Tr_{2...N}$ is even more complex than the partition function calculation.

\par The question arises: is it possible to use other methods (even approximate) to find the operators $\hat\varrho^{(1)}$ and $\hat\varrho^{(2)}$ directly without involvement of the higher-order operators? Indeed, such methods exist. First, the single-particle statistical operator $\hat\varrho^{(1)}$ separately or together with the operator $\hat\varrho^{(2)}$ for the equilibrium system, can be found from the variational principle taking into account properties of some thermodynamic quantities (for example, free energy) to take an extreme value when the many-particle system is in the statistical equilibrium state. Second, the statistical operator  $\hat\varrho^{(1)}$ can be found by solving the kinetic equation governing the non-equilibrium system evolution. Such equation was obtained in [5] under the assumption that the many-particle system evolution proceeds as a random Markov process. A variational method to find the equilibrium density matrices of the first and second orders was proposed in [6].  The method can be considered as a generalization of the Hartree-Fock-Slater method to the case of an arbitrary system of fermions which is in contact with a heat reservoir. In the present work, this method is used to describe the behavior of conduction electrons in metals.

\makeatletter\renewcommand{\@evenhead} {\vbox{\hbox
to\textwidth{\thepage\hfil\small\it New theory of superconductivity.
Method equilibrium density matrix \hfil}
\vskip 1mm\hrule\vskip -4mm}}

\makeatletter\renewcommand{\@oddhead} {\vbox{\hbox
to\textwidth{\hfill\small\it B.V. Bondarev \hfill\rm\thepage}
\vskip 1mm\hrule\vskip -4mm}}

\vskip 4mm
\centerline {\bf Statistical description of a fermion system }\vskip 2mm

\par The reduced description of the system consisting of $N$ identical particles (fermions) can be made by means of one- and two-particle density matrices
$$ \varrho_{11^\pr}=\varrho_{\alpha_1\alpha_1^\pr}\hh , \hskip 15mm
\varrho_{12,\h 1^\pr 2^\pr}=
\varrho_{\alpha_1\alpha_2,\h\alpha^\pr_1\alpha^\pr_2}\hh , \eqno (2) $$
where $\alpha$ is a system of quantum numbers describing the state of one particle. The single-particle density matrix $\varrho_{\alpha\alpha^\pr}$ satisfies the normalization condition
$$ \sum\limits_\alpha\varrho_{\alpha\alpha^\pr}=N\hh , $$
where $\varrho_{\alpha\alpha}$ is the probability of occupying the state $\alpha$.

\par The exact expression for the internal energy of the system of identical particles can be written using the density matrices (2) in the form [3]
$$ E=\sum\limits_{1,\h 1^\pr}\hh H_{11^\pr}\hs\varrho_{1^\pr 1}+
\frac{1}{2}\hh\sum\limits_{1,\h 2;\h 1^\pr, 2^\pr}\hh
H_{12,\h 1^\pr 2^\pr}\hs\varrho_{1^\pr 2^\pr,\h 12}
\hh . \eqno (3) $$
Here $H_{11^\pr}$ and $H_{12,\h 1^\pr 2^\pr}$ are matrix elements of the single-particle Hamiltonian $\hat H^{(1)}$ and the interaction Hamiltonian $\hat H^{(2)}$ of two particles, respectively:
$$ H_{11^\pr}=H_{\alpha_1\alpha^\pr_1}\hh , \hskip 15mm
H_{12\h ,1^\pr 2^\pr}=H_{\alpha_1\alpha_2,\h\alpha^\pr_1\alpha^\pr_2}
\hh . \eqno (4) $$

\par Taking into account that two-particle density matrix related to the system of fermions must be antisymmetric, we take the following approximate expression for the matrix:
$$ \varrho_{12,\h 1^\pr 2^\pr}=\varrho_{11^\pr}\hs\varrho_{22^\pr}- \varrho_{12^\pr}\hs\varrho_{21^\pr}\hh . \eqno (5) $$

\par Substitution of this expression into (3) gives the formula
$$ E=\sum\limits_{1,\h 1^\pr}\hh H_{11^\pr}\hs\varrho_{1^\pr 1}+
\sum\limits_{1,\h 2;\h 1^\pr, 2^\pr}\hh H_{12,\h 1^\pr 2^\pr}
\hs\varrho_{1^\pr 1}\hs\varrho_{2^\pr 2}\hh , \eqno (6) $$
which corresponds to the mean-field approximation.

\par The transition from the coordinate representation, in which the Hamiltonians are usually defined, to some $\alpha$-representation is made by means of the system of orthonormal wave functions $\varphi_\alpha(q)$, where $q\equiv\{{\bf r},\hh\sigma\}$; ${\bf r}$ is the particle's radius vector; $\sigma$ is a spin variable. By knowing these functions, the matrix elements of the Hamiltonians (4) can be calculated from the known equations
$$ H_{\alpha\alpha^\pr}=\int\hh\varphi^*_\alpha\hs\hat H^{(1)}\hh
\varphi_{\alpha\pr}\hs{\rm d}q\hh , \eqno (7) $$
$$ H_{12,\h 1^\pr 2^\pr}=\int\hh\Phi^*_{12}\hs\hat H^{(2)}\hh
\Phi_{1^\pr 2^\pr}\hs{\rm d}q_1\hs{\rm d}q_2\hh , \eqno (8) $$
where the integral sign represents integration over coordinates and summation over the spin variable; $\Phi_{12}$ is the Slater two-particle wave function:
$$ \Phi_{12}=\frac{1}{2}\hs\bigl\{\varphi_{\alpha_1}(q_1)\hh
\varphi_{\alpha_2}(q_2)-\varphi_{\alpha_1}(q_2)\hh
\varphi_{\alpha_2}(q_1)\bigr\}\hh . $$
By substituting this function into (8), we obtain the anti-symmetric matrix
$$ H_{12,\h 1^\pr 2^\pr}=\frac{1}{4}\hh(V_{12,\h 1^\pr 2^\pr}-V_{21,\h 1^\pr 2^\pr}-V_{12,\h 2^\pr 1^\pr}+V_{21,\h 2^\pr 1^\pr})
\hh , \eqno (9) $$
where
$$ V_{12,\h 1^\pr 2^\pr}=\int\hh\varphi^*_{\alpha_1}(q_1)\hs
\varphi^*_{\alpha_2}(q_2)\hs U(q_1,\h q_2)\hs\varphi_{\alpha^\pr_1}(q_1)
\hs\varphi_{\alpha^\pr_2}(q_2)\hs{\rm d}q_1\hs{\rm d}q_2\hh ; \eqno (10) $$
$U(q_1,\h q_2)$ is the potential energy of interaction between two fermions.

\par There exists a presentation in which the single-particle density matrix is diagonal, i.e. has the form
$$ \varrho_{nn^\pr}=w_n\hs\delta_{nn^\pr}\hh , \eqno (11) $$
where $n$ is the set of quantum numbers which determines the state of one particle in the new representation; $w_n$ -- diagonal elements of the density matrix; $\delta_{nn^\pr}$ -- Kronecker symbols. By definition, $w_n$ is the probability of occupation of the state $n$ by one of the particles. Therefore, the function $w_n$ describes the particle distribution over the states and satisfies the normalization condition
$$ \sum\limits_n\hs w_n=N\hh . \eqno (12) $$

\par The transition from the $n$-representation to the $\alpha$-representation in which the matrix elements (4) of the Hamiltonians $\hat H^{(1)}$ and $\hat H^{(2)}$ are defined is made through the unitary transformation
$$ \varrho_{\alpha\alpha^\pr}=\sum\limits_n\hs
\Psi_{\alpha n}\hs w_n\hs\Psi_{\alpha^\pr n}\hh , \eqno (13) $$
where $\Psi_{\alpha n}$ is a unitary matrix;
$$ \sum\limits_\alpha\hs\Psi^*_{\alpha n}\hs\Psi_{\alpha n^\pr}=
\delta_{nn^\pr}\hh . \eqno (14) $$

\par Using the distribution function $w_n$, it is possible to write down a well-known expression for the entropy of a system of fermions
$$ S=-\hs k_B\hh\sum\limits_n\bigl\{\hh w_n\hs\ln\h w_n+
(1-w_n)\hs\ln\h(1-w_n)\bigr\}\hh . \eqno (15) $$

\par Keeping in mind equations (6), (13) and (15), one can state that the free energy
$$ F=E-S\hs T $$
in the accepted approximation represents a functional depending on $w_n$ and $\Psi_{\alpha n}$. Since the equilibrium state of the system at fixed temperature and volume values corresponds to minimum of the free energy, the functions $w_n$ and $\Psi_{\alpha n}$ can be defined by minimization of the free energy taking into account conditions (12) and (14). Thus, we come to the conditional extreme problem. In order to solve this problem by Lagrange's method, let us compose the auxiliary functional
$$ \Omega=E-S\hs T-\mu\hh\sum\limits_n\h w_n-
\sum\limits_{n,\h n^\pr}\sum\limits_\alpha\Psi^*_{\alpha n}\hh \nu_{nn^\pr}\hh\Psi_{\alpha n^\pr}\hh , \eqno (16) $$
where $\mu$ and $\nu_{nn^\pr}$ are undetermined multipliers.

\par The functional extremum conditions (16) lead to the equations for the distribution function $w_n$ and matrix $\Psi_{\alpha n}$:
$$ \ln\hh\frac{1-w_n}{w_n}=\beta\hh(\varepsilon^{(1)}_n-\mu)
\hh ; \eqno (17) $$
$$ w_n\hh\sum\limits_{\alpha^\pr}\hh H^{(eff)}_{\alpha\alpha^\pr}
\hh\Psi_{\alpha^\pr n}-\sum\limits_{n^\pr}\hh\nu_{nn^\pr}\hh
\Psi_{\alpha n^\pr}\hh , \eqno (18) $$
where $\varepsilon^{(1)}_n$ is energy of one particle:
$$ \varepsilon^{(1)}_n=\varepsilon_n+\sum\limits_{n^\pr}\hh
V_{nn^\pr}\hs w_{n^\pr}\hh ; \hskip 15mm
\varepsilon_n=\sum\limits_{\alpha,\h\alpha^\pr}\hh
\Psi^*_{\alpha n}\hh H_{\alpha\alpha^\pr}\hh\Psi_{\alpha^\pr n}\hh ; $$
$$ V_{nn^\pr}=2\sum\limits_{1,\h 2;\h 1^\pr 2^\pr}
\Psi^*_{\alpha_1 n}\hh\Psi^*_{\alpha_2 n^\pr}\hh H_{12,\h 1^\pr 2^\pr}
\hh\Psi_{\alpha^\pr_1 n}\hh\Psi_{\alpha^\pr_2 n^\pr}\hh ; \hskip 15mm
V_{nn^\pr}=V_{n^\pr n}\hh ; $$
$H^{(eff)}_{\alpha\alpha^\pr}$ is the effective single-particle Hamiltonian in the mean-field approximation:
$$ H^{(eff)}_{\alpha\alpha^\pr}=H_{\alpha\alpha^\pr}+2
\sum\limits_{1,\h 1^\pr}\hh H_{\alpha\alpha_1,\h\alpha^\pr\alpha^\pr_1}
\hs\varrho_{\alpha^\pr_1\alpha_1}\hh . $$

\par The solution of the problem is significantly simplified in case that the considered properties of the system allow to predict in advance the representation, in which the density matrix must be diagonal. In this case, it only remains to solve equation (17). The solutions of this equation have some interesting features that are associated with its nonlinearity and nature of dependence of the core $V_{nn^\pr}$ on the quantum numbers $n$ and $n^\pr$. The aim of this work is to study these features and their physical consequences.

\vskip 4mm
\centerline{\bf Statistical description of electrons in a crystal lattice }\vskip 2mm

\par The atom arrangement in a crystal can be described by specifying the Bravais lattice and the atom positions in a particular unit cell. We will determine the position of one atom in the unit cell using a vector ${\bf R}$, and positions of all other atoms in the cell with respect to the first one by a vector ${\bf a}$. Assume that $s$ is a set of quantum numbers characterizing the wave function of one of the states of the electron localized in  vicinity of the atom, the position of which is determined by the vector ${\bf R}+{\bf a}$. Using the introduced symbols, we write down the orthonormal set of wave functions describing the localized electron states, in the form
$$ \varphi_\alpha(q)\equiv\varphi({\bf r}-{\bf R}-
{\bf a},\hh\sigma\hh|\hh{\bf a},\hh s)\hh , $$
where $\alpha=\{{\bf R},\hh{\bf a},\hh s\}$ -- set of quantum numbers defining the electron state in the crystal lattice. As these functions, it is convenient to use the Wannier functions. By means of these functions, it is possible to calculate the matrix elements of the Hamiltonians (7) and (8).

\par After that, the density matrix of the electron system equilibrium state in the crystal can be found by the method proposed in the previous section. Only some simplest types of Hamiltonians that simulate with some accuracy the interaction and behavior of the conduction electrons in real metals will be considered in this work.

\par Consider the cases where the unit cell has only one atom (${\bf a}=0$) and assume that the matrices (7) and (10) for valence electrons have the form
$$ H_{\alpha\alpha^\pr}=\varepsilon_{{\bf R}-{\bf R}^\pr}\hs
\delta_{ss^\pr}\hh ; \hskip 15mm V_{12,\h 1^\pr 2^\pr}=V_{{\bf R}_1{\bf R}_2,\h
{\bf R}^\pr_1{\bf R}^\pr_2}\hs\delta_{s_1s^\pr_1}\hs\delta_{s_2s^\pr_2}
\hh , \eqno (19) $$
where the $s$ parameter takes a finite number $G$ of different values;
$$ V_{{\bf R}_1{\bf R}_2,\h{\bf R}^\pr_1{\bf R}^\pr_2}=
\int\varphi({\bf r}_1)\hs\varphi({\bf r}_1+{\bf R}_1-{\bf R}^\pr_1)\hs U({\bf r}_1-{\bf r}_2+{\bf R}_1-{\bf R}_2)\hs\varphi({\bf r}_2)\hs
\varphi({\bf r}_2+{\bf R}_2-{\bf R}^\pr_2)\hs
{\rm d}{\bf r}_1\hs{\rm d}{\bf r}_2\hh ;  \eqno (20) $$
$\varphi({\bf r}-{\bf R})$ is the averaged wave function describing the electron localized in vicinity of the site ${\bf R}$;
$U({\bf r}_1-{\bf r}_2)$ -- the potential energy of Coulomb repulsion between two electrons. In this case of equilibrium, the density matrix describing the conduction electrons has the form
$$ \varrho_{\alpha\alpha^\pr}\equiv
\varrho^{(ss^\pr)}_{{\bf R}{\bf R}^\pr}=
\varrho_{{\bf R}{\bf R}^\pr}\hs\delta_{ss^\pr}\hh . \eqno (21) $$

\par Using (6), (9), (19) and (21), after simple transformations we obtain the following expression for the electron energy:
$$ E=G\hs\biggl\{\hs\sum\limits_{{\bf R},\h{\bf R}^\pr}
\varepsilon_{{\bf R}-{\bf R}^\pr}\hs\varrho_{\h{\bf R}^\pr{\bf R}}\hs +
\sum\limits_{\{\bf R\}}\hs H_{{\bf R}_1{\bf R}_2,\h{\bf R}^\pr_1{\bf R}^\pr_2}\hs
\varrho_{\h{\bf R}^\pr_1{\bf R}_1}\hs\varrho_{\h{\bf R}^\pr_2{\bf R}_2}\hs\biggr\}\hh , \eqno (22) $$
where
$\{\bf R\}={\bf R}_1,\hh{\bf R}_2,\hh{\bf R}^\pr_1,\hh{\bf R}^\pr_2$;
$$ H_{{\bf R}_1{\bf R}_2,\h{\bf R}^\pr_1{\bf R}^\pr_2}=
\frac{1}{4}\hs\bigl[\hh G\hs(
V_{{\bf R}_1{\bf R}_2,\h{\bf R}^\pr_1{\bf R}^\pr_2}+
V_{{\bf R}_2{\bf R}_1,\h{\bf R}^\pr_2{\bf R}^\pr_1})-
V_{{\bf R}_2{\bf R}_1,\h{\bf R}^\pr_1{\bf R}^\pr_2}-
V_{{\bf R}_1{\bf R}_2,\h{\bf R}^\pr_2{\bf R}^\pr_1}\bigr]
\hh . \eqno (23) $$

\par If electrons are distributed homogeneously over the crystal lattice sites, the density matrix $\varrho_{\h{\bf R}{\bf R}^\pr}$ can be expressed as
$$ \varrho_{\h{\bf R}{\bf R}^\pr}=\frac{1}{N_L}\hs\sum\limits_{\bf k}
\hs w_k\hs e^{\h{\bf{\rm i}\h k\h(R\h -\h R^\pr)}}\hs , \eqno (24) $$
where the summation is made over the ${\bf k}$ vectors belonging to the first Brillouin zone; $N_L$ -- number of sites in the lattice; $w_{\bf k}$ -- electron distribution function over wave vectors satisfying the normalization condition
$$ \frac{1}{N_L}\hs\sum\limits_{\bf k}\hh w_{\bf k}=\nu\hh ; \eqno (25) $$
$\nu$ -- the band filling degree: $\nu=N/G\hh N_L$.

\par Substitution of (24) into (22) gives
$$ E=G\hs\biggl(\sum\limits_{\bf k}\hh\varepsilon_{\bf k}\hs w_{\bf k}+
\frac{1}{2}\hs\sum\limits_{{\bf k},\h{\bf k}^\pr}\hh
V_{{\bf k}{\bf k}^\pr}\hs w_{\bf k}\hs w_{{\bf k}^\pr}\biggr)
\hh , \eqno (26) $$
where $\varepsilon_{\bf k}$ -- the electron kinetic energy:
$$ \varepsilon_{\bf k}=\sum\limits_{\bf R}\hh\varepsilon_{\bf R}\hs
e^{\h -\h{{\rm i}\h{\bf k}\h{\bf R}}}\hh ; $$
$V_{{\bf kk}^\pr}$ -- interaction energy of two electrons with the wave vectors
${\bf k}$ and ${\bf k}^\pr$;
$$ V_{{\bf kk}^\pr}=\frac{2}{N_L^2}\hs\sum\limits_{\{\bf R\}}\hh
H_{{\bf R}_1{\bf R}_2,\h{\bf R}^\pr_1{\bf R}^\pr_2}\hs
\exp\h\bigl[\hh{\rm i}\hh{\bf k}\hh
({\bf R}^\pr_1-{\bf R}_1)+{\rm i}\hh{\bf k}^{\h\pr}\hh
({\bf R}^\pr_2-{\bf R}_2)\bigr]\hh . \eqno (27) $$

\par Equation (24) is essentially a unitary transformation which diagonalizes the density matrix. In this case, (15) takes the form
$$ S=-\hs G\hs k_B\hh\sum\limits_{\bf k}\bigl\{\hh w_{\bf k}\hs
\ln\h w_{\bf k}+(1-w_{\bf k})\hs\ln\h(1-w_{\bf k})\bigr\}
\hh . $$

\par Minimizing the free energy and taking into account the normalization conditions (25), we obtain the integral equation that allows to find the distribution function $w_{\bf k}$ of conduction electrons over wave vectors
$$ \ln\hh\frac{1-w_{\bf k}}{w_{\bf k}}=
\beta\hh(\varepsilon^{(1)}_{\bf k}-\mu)\hh , \eqno  (28) $$
where $\varepsilon^{(1)}_{\bf k}$ -- the energy of one electron with the wave vector ${\bf k}$:
$$  \varepsilon^{(1)}_{\bf k}=\varepsilon_{\bf k}+
\sum\limits_{{\bf k}^\pr}\hh V_{{\bf kk}^\pr}\hs w_{{\bf k}^\pr}
\hh . \eqno (29) $$

\par In order to clarify the structure of the core (27) in the functionals (26) and (29), consider equation (20). Taking into account that among the matrix elements (20), the diagonal elements are the largest elements corresponding to
${\bf R}^\pr_1={\bf R}_1$ and ${\bf R}^\pr_2={\bf R}_2$, take the approximate formula
$$ V_{{\bf R}_1{\bf R}_2,\h{\bf R}^\pr_1{\bf R}^\pr_2}=
U_{{\bf R}_1-{\bf R}_2}\hs\delta_{{\bf R}_1{\bf R}^\pr_1}\hs
\delta_{{\bf R}_2{\bf R}^\pr_2}+ U^{(0)}_{{\bf R}_1-{\bf R}_2}\hs
\delta\h({\bf R}_1-{\bf R}^\pr_1-{\bf R}_2+{\bf R}^\pr_2)\hh , \eqno (30) $$
where $U_{{\bf R}_1-{\bf R}_2}$ is the average energy of the Coulomb interaction between two electrons localized at the sites ${\bf R}_1-{\bf R}_2$, and the second term approximates the non-diagonal elements. Strictly speaking, the function $U^{(0)}$ in (30) should depend not only on ${\bf R}_1-{\bf R}_2$, but also on ${\bf R}_1-{\bf R}^\pr_1$. Using equations(23),(26),(27) and (30), we obtain the following approximate expression for the electron interaction energy:
$$ E_{int}=\frac{1}{2}\hh G\hh\biggl(\nu\hs U_0\hh N-
\sum\limits_{{\bf k},\h{\bf k}^\pr}\hh I_{{\bf k}-{\bf k}^\pr}\hs
w_{\bf k}\hs w_{{\bf k}^\pr}+\sum\limits_{\bf k}\hh J_{\bf k}\hs w_{\bf k}\hs
w_{\h -\h{\bf k}}\biggr)\hh , \eqno (31) $$
where
$$ I_{{\bf k}-{\bf k}^\pr}=\frac{1}{N_L}\hh\sum\limits_{\bf R}\hh
U_{\bf R}\hs e^{\h{\rm i}\h({\bf k}-{\bf k}^\pr)\h{\bf R}}
\hh ; \eqno (32) $$
$$ J_{\bf k}=\sum\limits_{\bf R}\hh U^{(0)}_{\bf R}\hs
\Bigl(\h G-e^{\h -\h 2\h{\rm i}\h{\bf k\h R}}\Bigr)\hh . \eqno (33) $$

\par The first term in equation(31) is the energy of the direct Coulomb interaction between electrons which does not depend on the distribution function $w_{\bf k}$. The sum, following this term, is the electron exchange energy. The core
$I_{{\bf k}-{\bf k}^\pr}$ in this sum is a positive function which has the largest value at ${\bf k}^\pr=-\hh{\bf k}$ and decreases rapidly with increasing
$|{\bf k}^\pr-{\bf k}|$ because the Coulomb interaction is long-range. Since the exchange energy is negative, such behavior of the $I_{{\bf k}-{\bf k}^\pr}$ function causes the effective attraction between electrons with similar values of the wave vectors. Whereas positive terms in (31), containing the values $J_{\bf k}$, cause the effective repulsion between electrons with the wave vectors ${\bf k}$ and
$-\h{\bf k}$.

\par The single-electron energy (29) corresponding to the interaction energy (31), has the form
$$ \varepsilon^{(1)}_{\bf k}=\varepsilon_{\bf k}-
\sum\limits_{{\bf k}^\pr}\hh I_{{\bf k}-{\bf k}^\pr}\hs
w_{{\bf k}^\pr}+J_{\bf k}\hs w_{\h -\h{\bf k}}\hh . \eqno (34) $$

\par In the model of free electrons moving in the positive charge field of ions which are homogeneously distributed in space, the energy of one electron is described by equation [7]
$$  \varepsilon^{(1)}_{\bf k}=\frac{(\hbar\hh k)^2}{2\hh m}-
\frac{e^2}{2\hh\pi^2}\h\int\frac{w_{{\bf k}^\pr}\hs{\rm d}{\bf k}^\pr}
{|\hh{\bf k}^\pr-{\bf k}\hh|^2}\hh , \eqno (35) $$
where $m$ and $e$ are the electron mass and charge.

\par In both equation(34) and equation (35), the core $V_{{\bf kk}^\pr}$ has a common property. It has the lowest value at ${\bf k}^\pr={\bf k}$ and the highest value at ${\bf k}^\pr=-\hh{\bf k}$.

\par Unfortunately, using equations (34) or (35), it is not possible not only to analytically solve equation (28), but even to study it in any detail. Therefore, let us approximate the function (32) by the expression
$$ I_{{\bf k}-{\bf k}^\pr}=I\hs\delta_{{\bf kk}^\pr}\hh , $$
where $I$ is a positive constant and the value (33) is considered to be independent on the wave vector:
$$ J_{\bf k}=J>0\hh . $$
In this case, equation (34) has the form
$$ \varepsilon^{(1)}_{\bf k}=\varepsilon_{\bf k}-
I\hs w_{\bf k}+J\hs w_{\h -\h{\bf k}}\hh . \eqno (36) $$

\par Since the $V_{{\bf kk}^\pr}$ core's property described above is preserved in this approximation, it can be assumed that features of the distribution function $w_{\bf k}$ connected with this property will not be significantly changed.

\vskip 4mm
\centerline{\bf Electron distribution function over wave vectors }\vskip 2mm

\par Equation (36) gives the opportunity to transform the integral equation (28) into a system of two algebraic equations for two distribution function values $w_{\bf k}$ and $w_{\h -\h{\bf k}}$:
$$ \left.\begin{array}{l}
\ln\hh\ds\frac{1-w_{\h\bf k}}{w_{\h\bf k}}=
\beta\hh(E_{\bf k}-I\hs w_{\h\bf k}+J\hs w_{\h -\h{\bf k}})
\hs , \medskip \\
\ln\hh\ds\frac{1-w_{\h -\h\bf k}}{w_{\h -\h\bf k}}=
\beta\hh(E_{\bf k}-I\hs w_{\h -\h\bf k}+J\hs w_{\bf k})
\hs , \\ \end{array}\right\} \eqno (37) $$ where
$$ E_{\bf k}=\varepsilon_{\bf k}-\mu\hh . $$
It is seen from equations (37) that the distribution function $w_{\h\bf k}$ represents a complex function of the wave vector $\bf k$, in which the electron energy $E_{\bf k}$ is the intermediate variable:
$$ w_{\h\bf k}=w(E_{\bf k})\hh . \eqno (38) $$
The system of equations (37) has solutions of two types. The first type includes the symmetric distribution functions which for any wave vector values satisfy the condition
$$ w_{\h -\h\bf k}=w_{\h\bf k}\hh . \eqno (39) $$

\par It will be shown below that not all the solutions of system (37) have this property. There are distribution functions for which condition (39) breaks down for some values of the wave vector:
$$ w_{\h -\h\bf k}\neq w_{\h\bf k}\hh . \eqno (40) $$
Moreover, such anisotropic distribution function occurs even in absence of external fields.

\par Taking into account the normalization conditions (25), the average electron velocity can be defined as
$$ {\bf u}=\frac{G\hh\hbar}{m\hh N}\hs\sum\limits_{\bf k}\hh{\bf k}\hs w_{\bf k}\hh . \eqno (41) $$

\par If the distribution function is symmetrical, the average velocity (41) is equal to zero. For some anisotropic distribution functions, Equation (41) can give non-zero values of the electron ordered motion velocity, i.e. such distribution functions describe electric current. If the electric current states are stable and can exist even in the absence of external fields, such states of the electron system should be considered as superconducting.

\par The $I$ and $J$ parameters in equations (37) have a common origin and therefore are interrelated. Nevertheless, let us consider the solutions of the system (37) in three cases: 1) $I=0$; 2) $J=0$ и 3) $J=I$.

\par If $I=0$, then it is easy to exclude the $w_{\h -\h{\bf k}}$ value from the system (37) and obtain the equation
$$ \frac{1}{\beta}\hs\ln\hh\ds\frac{1-w_{\h\bf k}}{w_{\h\bf k}}=
E_{\bf k}+\frac{J}{1+\exp\h\bigl[\h\beta\h(E_{\bf k}+J\hh
w_{\bf k})\bigr]}\hh , \eqno (42) $$
The symmetric solutions of this equation are also solutions of the equation
$$ \ln\hh\ds\frac{1-w_{\h\bf k}}{w_{\h\bf k}}=
\beta\hh(E_{\bf k}+J\hs w_{\h{\bf k}})\hh , \eqno (43) $$
which is a consequence of system (37) and condition (39).

\par Fig. 1 shows the solutions of Equation (42) in the form of the probability $w$ dependence on the argument $x=E/J$ at different values of dimensionless temperature
$\theta=T/T_C$, where the critical temperature $T_C$ is defined as
$$ T_C=\frac{J}{4\hh k_B}\hh . $$

\newpage\phantom{x}

\vskip 4mm \unitlength=1mm \centerline{\begin{picture}(87,57)
\put(23.2,47.2){\it 1}\put(13.8,44.8){\it 2} \put(4.3,47.2){\it
3}\put(0,43){\it 4} \put(21,11){\it 1}\put(25.2,15.2){\it
2}\put(32,22){\it 3}\put(36,26){\it 4} \put(55,51){\it
1}\put(56,44){\it 2} \put(65,7){\it 1}\put(64.2,13){\it
2}\put(85,12.4){\it 3}\put(79,15){\it 4}
\put(0,10){\vector(1,0){87}}\put(79,5){$x$}
\multiput(20,10.5)(0,2){20}{\line(0,1){1.4}}
\put(20,10){\line(0,-1){1}}\put(17,5){$-1$}
\put(40,30){\circle*{0.7}}\multiput(40,10.5)(0,2){10}{\line(0,1){1.4}}
\put(40,10){\line(0,-1){1}}\put(35.9,5){$-\frac12$}
\put(60,10){\line(0,-1){1}}\put(59.2,5){0}
\put(60,10){\vector(0,1){50}}\put(62,58.5){$w(x,\hh\theta)$}
\put(0,50){\line(1,0){61}}\put(62,49){1,0}
\multiput(59.5,30)(-2,0){10}{\line(-1,0){1.4}}
\put(60,30){\line(1,0){1}}\put(62,29){0,5}
\put(60,10){\unitlength=1mm\special{em:linewidth 0.3pt} \put(-40,40)
{\special{em:moveto}} \put(0,0)  {\special{em:lineto}} }
\put(60,10){\unitlength=1mm\special{em:linewidth 0.3pt}
\put(11.62,0.3333)  {\special{em:moveto}} \put(9.528,0.6667)
{\special{em:lineto}} \put(8.160,1)  {\special{em:lineto}}
\put(5.360,2)  {\special{em:lineto}} \put(3.280,3)
{\special{em:lineto}} \put(1.492,4)  {\special{em:lineto}}
\put(-0.136,5)  {\special{em:lineto}} \put(-1.664,6)
{\special{em:lineto}} \put(-3.122,7)  {\special{em:lineto}}
\put(-4.536,8)  {\special{em:lineto}} \put(-5.908,9)
{\special{em:lineto}} \put(-7.252,10)  {\special{em:lineto}}
\put(-8.576,11)  {\special{em:lineto}} \put(-9.884,12)
{\special{em:lineto}} \put(-11.17,13)  {\special{em:lineto}}
\put(-12.45,14)  {\special{em:lineto}} \put(-13.72,15)
{\special{em:lineto}} \put(-14.99,16)  {\special{em:lineto}}
\put(-16.24,17)  {\special{em:lineto}} \put(-17.50,18)
{\special{em:lineto}} \put(-18.75,19)  {\special{em:lineto}}
\put(-20.00,20)  {\special{em:lineto}} \put(-21.25,21)
{\special{em:lineto}} \put(-22.50,22)  {\special{em:lineto}}
\put(-23.76,23)  {\special{em:lineto}} \put(-25.01,24)
{\special{em:lineto}} \put(-26.28,25)  {\special{em:lineto}}
\put(-27.55,26)  {\special{em:lineto}} \put(-28.83,27)
{\special{em:lineto}} \put(-30.12,28)  {\special{em:lineto}}
\put(-31.42,29)  {\special{em:lineto}} \put(-32.75,30)
{\special{em:lineto}} \put(-34.09,31)  {\special{em:lineto}}
\put(-35.46,32)  {\special{em:lineto}} \put(-36.88,33)
{\special{em:lineto}} \put(-38.34,34)  {\special{em:lineto}}
\put(-39.86,35)  {\special{em:lineto}} \put(-41.48,36)
{\special{em:lineto}} \put(-43.28,37)  {\special{em:lineto}}
\put(-45.36,38)  {\special{em:lineto}} \put(-48.16,39)
{\special{em:lineto}} \put(-49.52,39.33)  {\special{em:lineto}}
\put(-51.60,39.67)  {\special{em:lineto}} }
\put(60,10){\unitlength=1mm\special{em:linewidth 0.3pt}
\put(-43.84,36.59)  {\special{em:moveto}} \put(-43.76,35.77)
{\special{em:lineto}} \put(-43.60,34.84)  {\special{em:lineto}}
\put(-43.40,33.82)  {\special{em:lineto}} \put(-43.16,32.73)
{\special{em:lineto}} \put(-42.88,31.57)  {\special{em:lineto}}
\put(-40.96,23.96)  {\special{em:lineto}} \put(-40.64,22.64)
{\special{em:lineto}} \put(-40.32,21.32)  {\special{em:lineto}}
\put(-39.98,19.99)  {\special{em:lineto}} \put(-39.66,18.66)
{\special{em:lineto}} \put(-39.32,17.33)  {\special{em:lineto}}
\put(-38.98,16.00)  {\special{em:lineto}} \put(-38.64,14.66)
{\special{em:lineto}} \put(-38.26,13.33)  {\special{em:lineto}}
\put(-37.88,12.00)  {\special{em:lineto}} \put(-37.47,10.66)
{\special{em:lineto}} \put(-37.02,9.336)  {\special{em:lineto}}
\put(-36.54,8.000)  {\special{em:lineto}} \put(-35.98,6.668)
{\special{em:lineto}} \put(-35.32,5.336)  {\special{em:lineto}}
\put(-34.51,4.000)  {\special{em:lineto}} \put(-33.41,2.664)
{\special{em:lineto}} \put(-32.06,1.600)  {\special{em:lineto}}
\put(-31.57,1.336)  {\special{em:lineto}} \put(-30.27,0.800)
{\special{em:lineto}} \put(-29.20,0.536)  {\special{em:lineto}}
\put(-27.56,0.264)  {\special{em:lineto}} \put(-1.100,0.004)
{\special{em:lineto}} \put(-0.344,0.004)  {\special{em:lineto}}
\put(0.3148,0.020)  {\special{em:lineto}} \put(0.9626,0.044)
{\special{em:lineto}} \put(1.6024,0.096)  {\special{em:lineto}}
\put(2.2572,0.208)  {\special{em:lineto}} \put(2.8704,0.428)
{\special{em:lineto}} \put(3.4012,0.845)  {\special{em:lineto}}
\put(3.7720,1.567)  {\special{em:lineto}} \put(3.8710,3.401)
{\special{em:lineto}} \put(3.7714,4.232)  {\special{em:lineto}}
\put(3.6118,5.162)  {\special{em:lineto}} \put(3.4014,6.181)
{\special{em:lineto}} \put(3.1508,7.275)  {\special{em:lineto}}
\put(2.8701,8.431)  {\special{em:lineto}} \put(2.5689,9.634)
{\special{em:lineto}} \put(2.2545,10.88)  {\special{em:lineto}}
\put(1.9330,12.14)  {\special{em:lineto}} \put(1.6077,13.43)
{\special{em:lineto}} \put(1.2823,14.73)  {\special{em:lineto}}
\put(0.9579,16.04)  {\special{em:lineto}} \put(0.6335,17.36)
{\special{em:lineto}} \put(0.3093,18.69)  {\special{em:lineto}}
\put(-0.014,20.01)  {\special{em:lineto}} \put(-0.344,21.34)
{\special{em:lineto}} \put(-0.669,22.67)  {\special{em:lineto}}
\put(-1.016,24.00)  {\special{em:lineto}} \put(-1.401,25.34)
{\special{em:lineto}} \put(-1.729,26.67)  {\special{em:lineto}}
\put(-2.127,28.00)  {\special{em:lineto}} \put(-2.523,29.33)
{\special{em:lineto}} \put(-2.899,30.67)  {\special{em:lineto}}
\put(-3.456,32.00)  {\special{em:lineto}} \put(-3.993,33.33)
{\special{em:lineto}} \put(-4.607,34.67)  {\special{em:lineto}}
\put(-5.484,36.00)  {\special{em:lineto}} \put(-6.555,37.33)
{\special{em:lineto}} \put(-7.088,37.87)  {\special{em:lineto}}
\put(-8.390,38.67)  {\special{em:lineto}} \put(-9.074,38.93)
{\special{em:lineto}} \put(-10.68,39.47)  {\special{em:lineto}}
\put(-12.56,39.73)  {\special{em:lineto}} \put(-38.98,40.00)
{\special{em:lineto}} \put(-39.66,40.00)  {\special{em:lineto}}
\put(-40.32,39.98)  {\special{em:lineto}} \put(-40.96,39.96)
{\special{em:lineto}} \put(-41.60,39.90)  {\special{em:lineto}}
\put(-42.24,39.79)  {\special{em:lineto}} \put(-42.88,39.57)
{\special{em:lineto}} \put(-43.40,39.16)  {\special{em:lineto}}
\put(-43.76,38.43)  {\special{em:lineto}} \put(-43.84,36.59)
{\special{em:lineto}} }
\put(60,10){\unitlength=1mm\special{em:linewidth 0.3pt}
\put(27.00,1.2)  {\special{em:moveto}}
\put(25.99,1.3){\special{em:lineto}} \put(21.55,2)
{\special{em:lineto}} \put(17.10,3)  {\special{em:lineto}}
\put(13.58,4)  {\special{em:lineto}} \put(10.57,5)
{\special{em:lineto}} \put(7.880,6)  {\special{em:lineto}}
\put(5.408,7)  {\special{em:lineto}} \put(3.088,8)
{\special{em:lineto}} \put(0.8960,9)  {\special{em:lineto}}
\put(-1.208,10)  {\special{em:lineto}} \put(-3.244,11)
{\special{em:lineto}} \put(-5.224,12)  {\special{em:lineto}}
\put(-7.152,13)  {\special{em:lineto}} \put(-9.048,14)
{\special{em:lineto}} \put(-10.91,15)  {\special{em:lineto}}
\put(-12.76,16)  {\special{em:lineto}} \put(-14.58,17)
{\special{em:lineto}} \put(-16.40,18)  {\special{em:lineto}}
\put(-18.20,19)  {\special{em:lineto}} \put(-20.00,20)
{\special{em:lineto}} \put(-21.80,21)  {\special{em:lineto}}
\put(-23.60,22)  {\special{em:lineto}} \put(-25.42,23)
{\special{em:lineto}} \put(-27.24,24)  {\special{em:lineto}}
\put(-29.09,25)  {\special{em:lineto}} \put(-30.95,26)
{\special{em:lineto}} \put(-32.85,27)  {\special{em:lineto}}
\put(-34.78,28)  {\special{em:lineto}} \put(-36.76,29)
{\special{em:lineto}} \put(-38.79,30)  {\special{em:lineto}}
\put(-40.88,31)  {\special{em:lineto}} \put(-43.08,32)
{\special{em:lineto}} \put(-45.40,33)  {\special{em:lineto}}
\put(-47.88,34)  {\special{em:lineto}} \put(-50.56,35)
{\special{em:lineto}} \put(-53.56,36)  {\special{em:lineto}}
\put(-57.08,37)  {\special{em:lineto}} \put(-60.00,37.7)
{\special{em:lineto}} }
\put(60,10){\unitlength=1mm\special{em:linewidth 0.3pt}
\put(-36.61,28.25)  {\special{em:moveto}} \put(-36.56,27.54)
{\special{em:lineto}} \put(-36.49,26.81)  {\special{em:lineto}}
\put(-36.30,26.00)  {\special{em:lineto}} \put(-36.16,25.22)
{\special{em:lineto}} \put(-35.90,24.36)  {\special{em:lineto}}
\put(-35.60,23.46)  {\special{em:lineto}} \put(-35.30,22.58)
{\special{em:lineto}} \put(-34.92,21.62)  {\special{em:lineto}}
\put(-34.49,20.64)  {\special{em:lineto}} \put(-34.04,19.66)
{\special{em:lineto}} \put(-33.52,18.62)  {\special{em:lineto}}
\put(-32.93,17.55)  {\special{em:lineto}} \put(-32.27,16.47)
{\special{em:lineto}} \put(-31.54,15.34)  {\special{em:lineto}}
\put(-30.70,14.16)  {\special{em:lineto}} \put(-29.75,12.97)
{\special{em:lineto}} \put(-28.64,11.70)  {\special{em:lineto}}
\put(-27.28,10.36)  {\special{em:lineto}} \put(-25.57,8.904)
{\special{em:lineto}} \put(-23.82,7.696)  {\special{em:lineto}}
\put(-22.74,7.064)  {\special{em:lineto}} \put(-20.56,6.016)
{\special{em:lineto}} \put(-17.259,4.936)  {\special{em:lineto}}
\put(-13.507,4.348)  {\special{em:lineto}} \put(-11.362,4.300)
{\special{em:lineto}} \put(-9.7588,4.428)  {\special{em:lineto}}
\put(-8.4684,4.664)  {\special{em:lineto}} \put(-7.3924,4.988)
{\special{em:lineto}} \put(-6.4840,5.376)  {\special{em:lineto}}
\put(-5.7180,5.848)  {\special{em:lineto}} \put(-5.0736,6.376)
{\special{em:lineto}} \put(-4.5384,6.984)  {\special{em:lineto}}
\put(-4.1092,7.640)  {\special{em:lineto}} \put(-3.7788,8.388)
{\special{em:lineto}} \put(-3.5472,9.188)  {\special{em:lineto}}
\put(-3.4080,10.12)  {\special{em:lineto}} \put(-3.379, 11.74)
{\special{em:lineto}} \put(-3.4399,12.46)  {\special{em:lineto}}
\put(-3.5434,13.20)  {\special{em:lineto}} \put(-3.6889,13.99)
{\special{em:lineto}} \put(-3.8770,14.80)  {\special{em:lineto}}
\put(-4.1085,15.65)  {\special{em:lineto}} \put(-4.3842,16.53)
{\special{em:lineto}} \put(-4.7046,17.43)  {\special{em:lineto}}
\put(-5.0727,18.38)  {\special{em:lineto}} \put(-5.4901,19.35)
{\special{em:lineto}} \put(-5.9590,20.35)  {\special{em:lineto}}
\put(-6.4850,21.38)  {\special{em:lineto}} \put(-7.0731,22.44)
{\special{em:lineto}} \put(-7.7300,23.54)  {\special{em:lineto}}
\put(-8.4674,24.67)  {\special{em:lineto}} \put(-9.3005,25.83)
{\special{em:lineto}} \put(-10.252,27.04)  {\special{em:lineto}}
\put(-11.364,28.30)  {\special{em:lineto}} \put(-12.705,29.63)
{\special{em:lineto}} \put(-14.444,31.10)  {\special{em:lineto}}
\put(-15.347,31.76)  {\special{em:lineto}} \put(-16.176,32.30)
{\special{em:lineto}} \put(-17.259,32.94)  {\special{em:lineto}}
\put(-19.453,33.99)  {\special{em:lineto}} \put(-22.74,35.06)
{\special{em:lineto}} \put(-26.49,35.65)  {\special{em:lineto}}
\put(-28.64,35.70)  {\special{em:lineto}} \put(-30.24,35.57)
{\special{em:lineto}} \put(-31.54,35.34)  {\special{em:lineto}}
\put(-32.61,35.01)  {\special{em:lineto}} \put(-33.52,34.62)
{\special{em:lineto}} \put(-34.28,34.15)  {\special{em:lineto}}
\put(-34.92,33.62)  {\special{em:lineto}} \put(-35.45,33.02)
{\special{em:lineto}} \put(-35.90,32.36)  {\special{em:lineto}}
\put(-36.23,31.61)  {\special{em:lineto}} \put(-36.49,30.81)
{\special{em:lineto}} \put(-36.55,29.88)  {\special{em:lineto}}
\put(-36.61,28.25)  {\special{em:lineto}} }
\put(60,10){\unitlength=1mm\special{em:linewidth 0.3pt}
\put(33.81,1)  {\special{em:moveto}} \put(25.97,2)
{\special{em:moveto}} \put(20.86,3)  {\special{em:moveto}}
\put(16.87,4)  {\special{em:lineto}} \put(13.49,5)
{\special{em:lineto}} \put(10.48,6)  {\special{em:lineto}}
\put(7.736,7)  {\special{em:lineto}} \put(5.168,8)
{\special{em:lineto}} \put(2.752,9)  {\special{em:lineto}}
\put(0.4400,10)  {\special{em:lineto}} \put(-1.792,11)
{\special{em:lineto}} \put(-3.952,12)  {\special{em:lineto}}
\put(-6.056,13)  {\special{em:lineto}} \put(-8.120,14)
{\special{em:lineto}} \put(-10.14,15)  {\special{em:lineto}}
\put(-12.15,16)  {\special{em:lineto}} \put(-14.13,17)
{\special{em:lineto}} \put(-16.10,18)  {\special{em:lineto}}
\put(-18.05,19)  {\special{em:lineto}} \put(-20.00,20)
{\special{em:lineto}} \put(-21.95,21)  {\special{em:lineto}}
\put(-23.90,22)  {\special{em:lineto}} \put(-25.87,23)
{\special{em:lineto}} \put(-27.85,24)  {\special{em:lineto}}
\put(-29.85,25)  {\special{em:lineto}} \put(-31.88,26)
{\special{em:lineto}} \put(-33.94,27)  {\special{em:lineto}}
\put(-36.05,28)  {\special{em:lineto}} \put(-38.21,29)
{\special{em:lineto}} \put(-40.44,30)  {\special{em:lineto}}
\put(-42.76,31)  {\special{em:lineto}} \put(-45.16,32)
{\special{em:lineto}} \put(-47.72,33)  {\special{em:lineto}}
\put(-50.48,34)  {\special{em:lineto}} \put(-53.48,35)
{\special{em:lineto}} \put(-56.88,36)  {\special{em:lineto}}
\put(-60.00,36.7)  {\special{em:lineto}} }
\put(60,10){\unitlength=1mm\special{em:linewidth 0.3pt}
\put(-21.357,28.06)  {\special{em:moveto}} \put(-23.367,28.62)
{\special{em:lineto}} \put(-24.498,28.62)  {\special{em:lineto}}
\put(-25.346,28.60)  {\special{em:lineto}} \put(-26.022,28.47)
{\special{em:lineto}} \put(-26.574,28.24)  {\special{em:lineto}}
\put(-27.053,27.95)  {\special{em:lineto}} \put(-27.446,27.69)
{\special{em:lineto}} \put(-27.779,27.38)  {\special{em:lineto}}
\put(-28.059,27.04)  {\special{em:lineto}} \put(-28.299,26.66)
{\special{em:lineto}} \put(-28.460,26.28)  {\special{em:lineto}}
\put(-28.609,25.78)  {\special{em:lineto}} \put(-28.722,25.34)
{\special{em:lineto}} \put(-28.756,24.74)  {\special{em:lineto}}
\put(-28.756,23.74)  {\special{em:lineto}} \put(-28.722,23.34)
{\special{em:lineto}} \put(-28.609,22.78)  {\special{em:lineto}}
\put(-28.460,22.28)  {\special{em:lineto}} \put(-28.299,21.66)
{\special{em:lineto}} \put(-28.059,21.04)  {\special{em:lineto}}
\put(-27.779,20.38)  {\special{em:lineto}} \put(-27.446,19.69)
{\special{em:lineto}} \put(-27.053,18.95)  {\special{em:lineto}}
\put(-26.574,18.24)  {\special{em:lineto}} \put(-26.022,17.47)
{\special{em:lineto}} \put(-25.346,16.60)  {\special{em:lineto}}
\put(-24.498,15.62)  {\special{em:lineto}} \put(-23.367,14.62)
{\special{em:lineto}} \put(-21.357,13.06)  {\special{em:lineto}}
\put(-18.642,11.94)  {\special{em:lineto}} \put(-16.633,11.38)
{\special{em:lineto}} \put(-15.502,11.38)  {\special{em:lineto}}
\put(-14.653,11.40)  {\special{em:lineto}} \put(-13.979,11.54)
{\special{em:lineto}} \put(-13.427,11.76)  {\special{em:lineto}}
\put(-12.947,12.05)  {\special{em:lineto}} \put(-12.555,12.31)
{\special{em:lineto}} \put(-12.222,12.62)  {\special{em:lineto}}
\put(-11.941,12.96)  {\special{em:lineto}} \put(-11.700,13.34)
{\special{em:lineto}} \put(-11.541,13.72)  {\special{em:lineto}}
\put(-11.392,14.22)  {\special{em:lineto}} \put(-11.277,14.66)
{\special{em:lineto}} \put(-11.244,15.26)  {\special{em:lineto}}
\put(-11.244,16.26)  {\special{em:lineto}} \put(-11.277,16.66)
{\special{em:lineto}} \put(-11.392,17.22)  {\special{em:lineto}}
\put(-11.541,17.72)  {\special{em:lineto}} \put(-11.700,18.34)
{\special{em:lineto}} \put(-11.941,18.94)  {\special{em:lineto}}
\put(-12.222,19.62)  {\special{em:lineto}} \put(-12.555,20.31)
{\special{em:lineto}} \put(-12.947,21.05)  {\special{em:lineto}}
\put(-13.427,21.76)  {\special{em:lineto}} \put(-13.979,22.54)
{\special{em:lineto}} \put(-14.653,23.40)  {\special{em:lineto}}
\put(-15.502,24.38)  {\special{em:lineto}} \put(-16.633,25.38)
{\special{em:lineto}} \put(-18.642,26.94)  {\special{em:lineto}}
\put(-21.357,28.06)  {\special{em:lineto}} } \end{picture}}

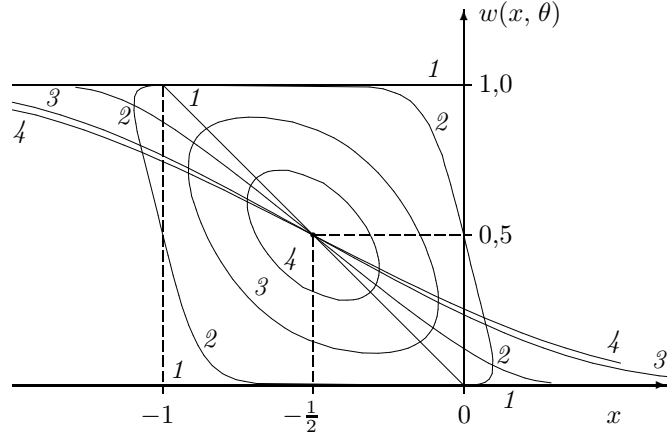
\vskip -2mm {\it Fig. 1. Dependence of distribution function $w$ on $x=(\varepsilon-\mu)/J$ at different values of dimensionless temperature $\theta=T/T_C$: 1 -- $\theta=0$; 2 -- $\theta=0.25$; 3 -- $\theta=0.8$; 4 -- $\theta=0.95$ }\vskip 2mm

\par At temperatures $T\geq T_C$, the distribution function is single-valued and satisfies the condition (39) for all values of $E$. At $T<T_C$, there exists a range of negative energy values $(E_1,\h E_2)$ at each point of which the function $w(E)$ can take any of three values:
$$ w^{(1)}(E)<w^{(0)}(E)<w^{(2)}(E)\hh . $$
Outside this range, the distribution function takes only one value $w^{(0)}(E)$. The function $w_{\h\bf k}=w^{(0)}(E_{\bf k})$ is a solution of equation (43) and describes the symmetrical electron distribution over wave vectors. At $T<T_C$, one more solution of equation (42) exists which can be written as
$$ w_{\h\bf k}=w^{(0)}(E_{\bf k})\hskip 5mm \hbox{at} \hskip 5mm
E_{\bf k}\leq E_1\h , \hs E_{\bf k}\geq E_2\hh . $$
If
$$ E_1<E_{\bf k}<E_2\hh , $$
then
$$ w_{\h\bf k}=w^{(1)}(E_{\bf k})\hh , \hskip 7mm
w_{\h -\h{\bf k}}=w^{(2)}(E_{\bf k})\hh \eqno (44 ) $$
or
$$ w_{\h\bf k}=w^{(2)}(E_{\bf k})\hh , \hskip 7mm
w_{\h -\h\bf k}=w^{(1)}(E_{\bf k})\hh . $$

\par At $T=0$, the symmetric solution of equation (42) has the form
$$ w_{\h\bf k}=\left\{\begin{array}{ccc}
1 & \hbox{at} & \varepsilon_{\bf k}\leq \mu-J
\hs , \medskip \\
-\hh\ds\frac{1}{J}\hh(\varepsilon_{\bf k}-\mu) & \hbox{at} &
\mu-J<\varepsilon_{\bf k}<\mu\hs , \medskip \\
0 & \hbox{at} & \varepsilon_{\bf k}\geq\mu
\hs , \\ \end{array}\right. \eqno (45) $$
and the anisotropic solution (44) is that
$$ w_{\hh\bf k}=1 \hskip 7mm \hbox{at} \hskip 7mm \varepsilon_{\bf k}\leq \mu-J\hh , $$
$$ w_{\hh\bf k}=1\hh , \hskip 7mm w_{\h -\h\bf k}=0 \hskip 7mm \hbox{at} \hskip 7mm
w_{\hh\bf k}=0\hh , \hskip 7mm w_{\h -\h\bf k}=1 \eqno (46) $$
$$ \hbox{at} \hskip 7mm \mu-J<\varepsilon_{\bf k}<\mu\hh , \eqno (47) $$
$$ w_{\hh\bf k}=0 \hskip 7mm \hbox{at} \hskip 7mm \varepsilon_{\bf k}\geq\mu\hh . $$

\par It is seen from equations (46), that there is a layer S under the Fermi surface, defined by inequations (47), in which the distribution function is anisotropic: among two states with the wave vectors $\bf k$ and $-\h{\bf k}$ in this layer, one state is free when the other one is certainly occupied.

\par If the electron gas state is described by the anisotropic distribution function (44) or (46), the electron ordered motion velocity  (41) can have any value in the range from zero to some $u_m$. The average velocity is equal to zero when the pairs of free and occupied states with the wave vectors $\bf k$ and $-\h{\bf k}$ are distributed chaotically in the layer $S$. When in one half of this layer (for example, with $k_x>0$) all the states are occupied and in the second one (with $k_x<0$) are free, the electron ordered motion velocity is maximized. The electron ordered motion velocity value is determined by the nature of initial state of the electron gas. When the anisotropic solution is stable with respect to small changes in external conditions, this electron velocity value will persist indefinitely. It means that the metal has become a superconductor.

\par When performing calculations, the summation in equations (25), (26) and (41) can be conveniently replaced by integration:
$$ \frac{G\hh V}{(2\hh\pi)^3}\hs\int w_{\hh\bf k}\hs{\rm d}{\bf k}
=N\hh , \eqno (48) $$
$$ E=\frac{G\hh V}{(2\hh\pi)^3}\hs\int\biggl(\varepsilon_{\bf k}+
\frac{1}{2}\hh\bigr(-\hh I\hs w_{\hh\bf k}+J\hs w_{\h -\h\bf k}\bigl)
\biggr)\hh w_{\hh\bf k}\hs{\rm d}{\bf k}\hh , \eqno (49) $$
$$ {\bf u}=\frac{G\hh\hbar\hh V}{(2\hh\pi)^3\hh m\hh N}\hs{\bf k}\hs\int
w_{\hh\bf k}\hs{\rm d}{\bf k}\hh , \eqno (50) $$
where $V$ is the volume of the crystal. Expression (49) for the electron internal energy corresponds to equation (36) for the internal energy of a single electron.

\par Assume the approximation formula for the electron kinetic energy
$$ \varepsilon_{\bf k}=\frac{(\hbar\hh k)^2}{2\hh m}\hh . \eqno (51) $$

\par Using equations (49) and (51), it is possible to show that the electron energy $E^{(s)}$ in case of the anisotropic distribution is lower then that in case of the symmetrical distribution over wave vectors. For example, when $Т=0$ we have
$$ E^{(s)}-E^{(0)}=-\hs\frac{G\hs V\hh m\hs p_F\hh J^2}
{24\hh\pi^2\hh\hbar^3}\hh , \eqno (52) $$
where $p_F$ is the Fermi momentum:
$$ p_F=\sqrt{2\hh m\hh\mu_{\rm o}}\hs , $$
$$ \mu_{\rm o}=\frac{\hbar^2}{2\hh m}\hh\biggl(\frac{6\hh\pi^2\hh N}
{G\hs V}\biggr)^{2/3}\hh . $$

\par The negative sign of the difference (52) means that the electron state described by the anisotropic distribution function (44) or (46) is stable. In other words, this state is superconducting. Note that in the considered model, the superconductivity is caused by repulsion between electrons with the wave vectors $\bf k$ and $-\hh\bf k$.

\par As a parameter characterizing the superconducting state of the electron system, it is possible to take the quantity
$$ \xi=\max\hh|\hh w^{(2)}-w^{(1)}\hh|\hh , \eqno (53) $$
which is the measure of the distribution function anisotropy. This quantity has the largest value $(\xi_m=1)$ at $T=0$ and decreases monotonically with increasing temperature, approaching to zero at $T\geq T_C$.

\par Using equation (50) it is possible to show that at $T=0$ the highest electron ordered motion velocity value is
$$ u_m=\frac{3\hh J}{4\hs p_F}\hh . $$

\par According to equations (36) and (46), the electron energy at $T=0$ is
$$ \varepsilon^{(1)}_{\h\bf k}=\left\{\begin{array}{ccc}
\varepsilon_{\hh\bf k}+J & \hbox{at} & \varepsilon_{\bf k}\leq \mu-J
\hs , \medskip \\
\varepsilon_{\hh\bf k} & \hbox{at} & \varepsilon_{\bf k}>\mu-J
\hs . \\ \end{array}\right. $$
It is seen from these equations that the gap in the conduction electron energy spectrum is absent.

\par Assume that $J=0$. In this case, the system (37) splits into two identical equations of the form
$$ \ln\hh\ds\frac{1-w_{\h\bf k}}{w_{\h\bf k}}=
\beta\hh(E_{\bf k}-I\hs w_{\h\bf k})\hh . \eqno (54) $$
In this case, the distribution function satisfies the condition (39) for all values of $\bf k$.

\par The plots of functions that are solutions of equation (54) are shown in Fig. 2. At temperatures
$$ T\geq T_C=\frac{I}{4\hh k_B} $$
the function $w=w(E)$ decreases monotonically. At $T<T_C$, at each point of the energy values range $(E_1,\hh E_2)$ existing in the area of positive $E_{\bf k}$, the distribution functions can have any of the three possible values:
$$ w^{(1)}(E)<w^{(0)}(E)<w^{(2)}(E)\h . $$
Outside this range, the distribution function has only one value $w^{(0)}(E)$. The electron gas energy in the state described by the distribution function
$$ w_{\h\bf k}=\left\{\begin{array}{ccc}
w^{(0)}(E) & \hbox{at} & E\leq E_1\hh , \hskip 2mm
E\geq E_2\hh , \medskip \\
w^{(1)}(E) & \hbox{at} & E_1<E<E_2
\hh , \\ \end{array}\right. \eqno (55) $$
takes the smallest value. At $E=E_1$, the function (55) experiences a discontinuity, as a result of which, a gap arises in the conduction electron energy spectrum.

\par At $T=0$, the function (55) takes the form
$$ w_{\h\bf k}=\left\{\begin{array}{ccc}
1 & \hbox{at} & \varepsilon_{\bf k}\leq\mu\hh , \medskip \\
0 & \hbox{at} & \varepsilon_{\bf k}>\mu\hh . \\ \end{array}\right. $$
Here, the energy of one electron is
$$ \varepsilon_{\bf k}^{(1)}=\left\{\begin{array}{ccc}
\varepsilon_{\bf k}-I & \hbox{at} & \varepsilon_{\bf k}<\mu\hh , \medskip \\
\varepsilon_{\bf k} & \hbox{at} & \varepsilon_{\bf k}>\mu\hh . \\ \end{array}\right. $$
It is seen from the equation analysis that the electron energy spectrum has a gap with width $I$. It should be noted that the gap formation is caused by the attraction between electrons with the wave vectors $\bf k$ и $\bf k^\pr$ which are equal to each other.

\vskip 10mm \unitlength=1mm \centerline{\begin{picture}(85,57)
\put(24,17){\it 1}\put(32,17){\it 2}\put(43,17){\it
3}\put(61,17){\it 4} \put(0,10){\vector(1,0){85}}
\put(65.5,5){$x$\hs=\hs$(\varepsilon$\hs--\hs$\mu)/I$}
\put(20,10){\line(0,-1){1}}\put(19.2,5){0}
\put(40,10){\line(0,-1){1}}\put(37.7,5){0,5}
\put(60,10){\line(0,-1){1}}\put(59.2,5){1}
\put(20,10){\vector(0,1){50}}\put(22,58){$w(x,\hh\theta)$}
\put(14,11){$0$} \put(20,30){\line(1,0){1}}\put(14,29){$0,5$}
\put(14,51.5){$1,0$}
\multiput(60,10.5)(0,2.013){20}{\line(0,1){1.4}}
\put(20,10){\unitlength=1mm\special{em:linewidth 0.3pt}
\put(-20,40){\special{em:moveto}} \put(40,40) {\special{em:lineto}}
\put(0,0)   {\special{em:lineto}} }
\put(20,10){\unitlength=1mm\special{em:linewidth 0.3pt}
\put(15.27,39.7){\special{em:moveto}}
\put(17.65,39.5){\special{em:lineto}}
\put(19.16,39.3){\special{em:lineto}} \put(20.68,39)
{\special{em:lineto}} \put(22.01,38.6){\special{em:lineto}}
\put(23.28,38)  {\special{em:lineto}} \put(24.44,37)
{\special{em:lineto}} \put(25.02,36)  {\special{em:lineto}}
\put(25.27,35)  {\special{em:lineto}} \put(25.33,34)
{\special{em:lineto}} \put(25.24,33)  {\special{em:lineto}}
\put(25.07,32)  {\special{em:lineto}} \put(24.82,31)
{\special{em:lineto}} \put(24.50,30)  {\special{em:lineto}}
\put(23.76,28)  {\special{em:lineto}} \put(22.90,26)
{\special{em:lineto}} \put(21.97,24)  {\special{em:lineto}}
\put(21.00,22)  {\special{em:lineto}} \put(20.00,20)
{\special{em:lineto}} \put(19.00,18)  {\special{em:lineto}}
\put(18.03,16)  {\special{em:lineto}} \put(17.10,14)
{\special{em:lineto}} \put(16.24,12)  {\special{em:lineto}}
\put(15.50,10)  {\special{em:lineto}} \put(15.18,9)
{\special{em:lineto}} \put(14.93,8)   {\special{em:lineto}}
\put(14.76,7)   {\special{em:lineto}} \put(14.68,6)
{\special{em:lineto}} \put(14.73,5)   {\special{em:lineto}}
\put(14.98,4)   {\special{em:lineto}} \put(15.56,3)
{\special{em:lineto}} \put(16.72,2)   {\special{em:lineto}}
\put(17.86,1.45){\special{em:lineto}} \put(19.32,1.0)
{\special{em:lineto}} \put(20.84,0.7) {\special{em:lineto}}
\put(22.35,0.5) {\special{em:lineto}} }
\put(20,10){\unitlength=1mm\special{em:linewidth 0.3pt}
\put(-4.18,39.5)  {\special{em:moveto}} \put(2.36,39)
{\special{em:lineto}} \put(8.56,38)  {\special{em:lineto}}
\put(11.88,37)  {\special{em:lineto}} \put(14.03,36)
{\special{em:lineto}} \put(15.54,35)  {\special{em:lineto}}
\put(16.66,34)  {\special{em:lineto}} \put(18.14,32)
{\special{em:lineto}} \put(19.01,30)  {\special{em:lineto}}
\put(19.53,28)  {\special{em:lineto}} \put(19.81,26)
{\special{em:lineto}} \put(19.94,24)  {\special{em:lineto}}
\put(20.00,22)  {\special{em:lineto}} \put(20.00,20)
{\special{em:lineto}} \put(20.00,18)  {\special{em:lineto}}
\put(20.06,16)  {\special{em:lineto}} \put(20.19,14)
{\special{em:lineto}} \put(20.47,12)  {\special{em:lineto}}
\put(20.99,10)  {\special{em:lineto}} \put(21.60,8.5)
{\special{em:lineto}} \put(21.86,8.0)  {\special{em:lineto}}
\put(22.16,7.5)  {\special{em:lineto}} \put(22.51,7.0)
{\special{em:lineto}} \put(22.90,6.5)  {\special{em:lineto}}
\put(23.35,6.0)  {\special{em:lineto}} \put(23.86,5.5)
{\special{em:lineto}} \put(24.46,5.0)  {\special{em:lineto}}
\put(25.15,4.5)  {\special{em:lineto}} \put(25.97,4.0)
{\special{em:lineto}} \put(26.95,3.5)  {\special{em:lineto}}
\put(28.12,3.0)  {\special{em:lineto}} \put(29.58,2.5)
{\special{em:lineto}} \put(31.44,2.0)  {\special{em:lineto}}
\put(33.95,1.5)  {\special{em:lineto}} \put(37.64,1.0)
{\special{em:lineto}} \put(44.20,0.5)  {\special{em:lineto}} }
 \put(20,10){\unitlength=1mm\special{em:linewidth
0.3pt} \put(-20.00,37.9)  {\special{em:moveto}} \put(-16.66,37.5)
{\special{em:lineto}} \put(-13.24,37)  {\special{em:lineto}}
\put(-10.42,36.5)  {\special{em:lineto}} \put(-7.96,36)
{\special{em:lineto}} \put(-5.82,35.5)  {\special{em:lineto}}
\put(-3.92,35)  {\special{em:lineto}} \put(-2.22,34.5)
{\special{em:lineto}} \put(-0.68,34)  {\special{em:lineto}}
\put(0.70,33.5)  {\special{em:lineto}} \put(1.98,33)
{\special{em:lineto}} \put(3.18,32.5)  {\special{em:lineto}}
\put(4.28,32)  {\special{em:lineto}} \put(8.02,30)
{\special{em:lineto}} \put(11.06,28)  {\special{em:lineto}}
\put(13.62,26)  {\special{em:lineto}} \put(15.89,24)
{\special{em:lineto}} \put(17.99,22)  {\special{em:lineto}}
\put(20.00,20)  {\special{em:lineto}} \put(22.01,18)
{\special{em:lineto}} \put(24.11,16)  {\special{em:lineto}}
\put(26.38,14)  {\special{em:lineto}} \put(28.94,12)
{\special{em:lineto}} \put(31.98,10)  {\special{em:lineto}}
\put(35.72,8)  {\special{em:lineto}} \put(38.02,7)
{\special{em:lineto}} \put(39.30,6.5)  {\special{em:lineto}}
\put(40.72,6)  {\special{em:lineto}} \put(42.24,5.5)
{\special{em:lineto}} \put(43.92,5)  {\special{em:lineto}}
\put(45.84,4.5)  {\special{em:lineto}} \put(47.96,4)
{\special{em:lineto}} \put(50.44,3.5)  {\special{em:lineto}}
\put(53.24,3)  {\special{em:lineto}} \put(56.68,2.5)
{\special{em:lineto}} \put(60.88,2)  {\special{em:lineto}} }
\end{picture}}

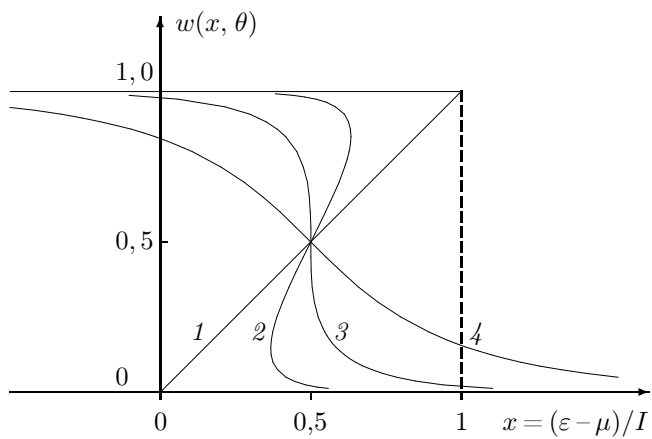
\vskip -2mm\centerline {\it Fig. 2. The distribution function $w(x,\hh\theta)$ of electrons over energy at different temperature values $\theta=T/T_C$: }
\centerline {\it 1 -- $\theta=0$; 2 -- $\theta=0.5$; 3 -- $\theta=1$; 4 -- $\theta=2$ } \vskip 4mm

\par Finally, consider the case where the $I$ and $J$ parameters are equal. In this case, both symmetric and anisotropic distribution functions will be among the solutions of system (37). The solution of system (37) satisfying condition (39) is a Fermi-Dirac function.

\newpage

\par One of the anisotropic distribution functions at $T=0$ is such that
$$ w_{\hh\bf k}=1 \hskip 7mm \hbox{at} \hskip 7mm
\varepsilon_{\bf k}\leq \mu-J\hh , $$
$$ w_{\hh\bf k}=1\hh , \hskip 7mm w_{\h -\h\bf k}=0 \hskip 7mm
\hbox{or} \hskip 7mm
w_{\hh\bf k}=0\hh , \hskip 7mm w_{\h -\h\bf k}=1 \hskip 7mm
\hbox{at} \hskip 7mm \mu-J<\varepsilon_{\bf k}<\mu\hh , \eqno (56) $$
$$ w_{\hh\bf k}=0 \hskip 7mm \hbox{at} \hskip 7mm \varepsilon_{\bf k}\geq\mu\hh . $$

\par By using these equations, we find that the one-electron energy (36) is
$$ \varepsilon_{\bf k}^{(1)}=\left\{\begin{array}{ccc}
\varepsilon_{\bf k} & \hbox{at} & \varepsilon_{\bf k}\leq\mu-I
\hh , \medskip \\
\varepsilon_{\bf k}-I & \hbox{at} & \mu-I<\varepsilon_{\bf k}<\mu
\hh , \medskip \\
\varepsilon_{\bf k} & \hbox{at} & \varepsilon_{\bf k}\geq\mu\hh . \\ \end{array}\right. \eqno (57) $$

\par The analysis of equations (56) and (57) shows that there is a layer $S$ below the Fermi surface: $\mu-I<\varepsilon_{\bf k}<\mu$, where the repulsion between electrons with the wave vectors $\bf k$ and $-\h{\bf k}$ makes them superconducting, and the distribution function has discontinuity on the Fermi surface. In this case, the electron energy spectrum has a gap with width $I$.

\par Using equation (49), it is easy to calculate the conduction electron energies at $T=0$ in the states described by different equilibrium distribution functions. The lowest energy corresponds to the anisotropic distribution function (56). Moreover, this energy does not depend on whether the electronic system state (56) is a current state or not. The energy $E^{(s)}$ differs from the energy $E^{(n)}$ of electrons in the normal state which is described by the Fermi-Dirac function, to the amount of
$$ E^{(s)}-E^{(n)}=-\hh\frac{G\hs V\hh m\hs p_F\hh I^2}
{16\hh\pi^2\hh\hbar^3}\hh . \eqno (58) $$

\par Fig. 3 shows the probability $w$ dependence on the argument $x=(\varepsilon-\mu)/I$ for the case $J=I$ at different values of the dimensionless temperature $\theta=T/T_C^{\hh\pr}$, where
$$ T^{\hh\pr}_C=\frac{I}{2\hh k_B}\hh . $$

\par At temperatures
$$ T<T_C=\frac{1}{\sqrt 3}\hs T^{\hh\pr}_C $$
the function $w=w(E)$ is single-valued only outside some interval
$[\hh E^{\hh\pr}_1,\hh E^{\hh\pr}_2\h]$ on the $E$-axis and coincides with the Fermi - Dirac function $w=w_F(E)$. In the intervals $[\hh E^{\hh\pr}_1,\hh E_1\h)$ and
$(\hh E_2,\hh E^{\hh\pr}_2\h]$, the distribution function can have any of the values
$$ w^{(1)}(E)<w^{(1^\pr)}(E)<w_F(E)<w^{(2^\pr)}(E)<w^{(2)}(E)\hh , $$
and in the interval $[\hh E_1,\hs E_2\hh]$ -- one of the three values
$$ w^{(1)}(E)<w_F(E)<w^{(2)}(E)\hh . $$
It can be shown that the electron system has the lowest internal energy when its state is described by the following anisotropic distribution function:
$$ \left.\begin{array}{ccc}
w_{\hh\bf k}=w_F(E) & \hbox{at} & E<E_1\hh , \hskip 3mm \hs E>E_2\hh ;
\\ \end{array}\right. $$
$$ \left.\begin{array}{cccc}
w_{\hh\bf k}=w^{(2)}(E)\h , &  w_{\h -\h\bf k}=w^{(1)}(E) & \hbox{at} &
E_1\leq E\leq E_2 \hh . \\ \end{array}\right. \eqno (59) $$

\par At temperatures $T_C<T<T^\pr_C$, the equalities $E^\pr_C=E_1$ and $E^\pr_2=E_2$ are fulfilled and the anisotropic distribution function has the form (44). At
$T\geq T^\pr_C$, the Fermi - Dirac function is the unique solution of equations (37).

\newpage\phantom{x}

\vskip 4mm \unitlength=1mm
\centerline{\begin{picture}(82,57)\put(-2,59){\it a\hs{\rm )}}
\put(-2,10){\vector(1,0){87}}\put(77,5){$\varepsilon-\mu$}
\put(20,10){\vector(0,1){51}}\put(22,59.5){$w$}\put(17,51.5){1}
\put(0,10){\line(0,-1){1}}\put(-2,5){$-\hh I$}
\put(80,10){\line(0,-1){1}}\put(79,12){$J$}
\put(20,10){\line(0,-1){1}}\put(19.3,5){$0$}
\multiput(60,11)(0,1.98){20}{\line(0,1){1.4}}
\put(60,10){\line(0,-1){1}}\put(55.9,5){$J-I$}
\multiput(-2,50)(2.02,0){31}{\line(1,0){1.4}}
\put(40,30){\circle*{0.7}}
\put(29,33){$Z$}\put(48.5,25){$Z$}
\put(20,10){\unitlength=1mm\special{em:linewidth 0.3pt}
\put(20.00,0.9)  {\special{em:moveto}} \put(18.50,1.0)
{\special{em:lineto}} \put(17.00,1.2)  {\special{em:lineto}}
\put(15.02,1.489)  {\special{em:lineto}} \put(12.69,1.990)
{\special{em:lineto}} \put(11.18,2.442)  {\special{em:lineto}}
\put(10.06,2.874)  {\special{em:lineto}} \put(9.200,3.295)
{\special{em:lineto}} \put(8.512,3.710)  {\special{em:lineto}}
\put(7.514,4.528)  {\special{em:lineto}} \put(7.148,4.932)
{\special{em:lineto}} \put(6.850,5.336)  {\special{em:lineto}}
\put(6.610,5.736)  {\special{em:lineto}} \put(6.418,6.136)
{\special{em:lineto}} \put(6.157,6.932)  {\special{em:lineto}}
\put(6.024,7.724)  {\special{em:lineto}} \put(6.093,9.692)
{\special{em:lineto}} \put(6.545,11.64)  {\special{em:lineto}}
\put(7.236,13.58)  {\special{em:lineto}} \put(8.080,15.48)
{\special{em:lineto}} \put(9.016,17.37)  {\special{em:lineto}}
\put(10.00,19.22)  {\special{em:lineto}} \put(11.00,21.04)
{\special{em:lineto}} \put(11.98,22.81)  {\special{em:lineto}}
\put(12.92,24.52)  {\special{em:lineto}} \put(13.79,26.17)
{\special{em:lineto}} \put(14.58,27.74)  {\special{em:lineto}}
\put(15.25,29.23)  {\special{em:lineto}} \put(15.79,30.62)
{\special{em:lineto}} \put(16.18,31.91)  {\special{em:lineto}}
\put(16.43,33.08)  {\special{em:lineto}} \put(16.43,34.00)
{\special{em:lineto}} }
\put(20,10){\unitlength=1mm\special{em:linewidth 0.3pt}
\put(16.43,34.00)  {\special{em:moveto}} \put(16.43,35.08)
{\special{em:lineto}} \put(16.43,35.08)  {\special{em:lineto}}
\put(16.18,35.91)  {\special{em:lineto}} \put(15.79,36.62)
{\special{em:lineto}} \put(15.25,37.23)  {\special{em:lineto}}
\put(14.58,37.74)  {\special{em:lineto}} \put(13.79,38.17)
{\special{em:lineto}} \put(12.92,38.52)  {\special{em:lineto}}
\put(11.98,38.81)  {\special{em:lineto}} \put(11.00,39.04)
{\special{em:lineto}} \put(10.00,39.22)  {\special{em:lineto}}
\put(9.016,39.37)  {\special{em:lineto}} \put(8.080,39.48)
{\special{em:lineto}} \put(7.236,39.58)  {\special{em:lineto}}
\put(6.545,39.64)  {\special{em:lineto}} \put(6.093,39.69)
{\special{em:lineto}} \put(5.993,39.71)  {\special{em:lineto}}
\put(6.157,39.73)  {\special{em:lineto}} \put(6.610,39.74)
{\special{em:lineto}} \put(7.148,39.73)  {\special{em:lineto}}
\put(8.512,39.71)  {\special{em:lineto}} \put(9.200,39.70)
{\special{em:lineto}} \put(10.06,39.67)  {\special{em:lineto}}
\put(11.18,39.64)  {\special{em:lineto}} \put(12.69,39.59)
{\special{em:lineto}} \put(15.02,39.49)  {\special{em:lineto}}
\put(20.00,39.15)  {\special{em:lineto}} }
\put(20,10){\unitlength=1mm\special{em:linewidth 0.3pt}
\put(20.00,39.15)  {\special{em:moveto}} \put(22.50,38.90)
{\special{em:lineto}} \put(24.98,38.51)  {\special{em:lineto}}
\put(27.31,38.01)  {\special{em:lineto}} \put(28.82,37.56)
{\special{em:lineto}} \put(29.94,37.13)  {\special{em:lineto}}
\put(30.80,36.70)  {\special{em:lineto}} \put(31.49,36.29)
{\special{em:lineto}} \put(32.04,35.88)  {\special{em:lineto}}
\put(32.49,35.47)  {\special{em:lineto}} \put(32.86,35.07)
{\special{em:lineto}} \put(33.15,34.66)  {\special{em:lineto}}
\put(33.39,34.26)  {\special{em:lineto}} \put(33.58,33.86)
{\special{em:lineto}} \put(33.84,33.07)  {\special{em:lineto}}
\put(33.98,32.28)  {\special{em:lineto}} \put(34.01,31.49)
{\special{em:lineto}} \put(33.96,30.70)  {\special{em:lineto}}
\put(33.84,29.92)  {\special{em:lineto}} \put(33.67,29.14)
{\special{em:lineto}} \put(33.46,28.36)  {\special{em:lineto}}
\put(32.77,26.42)  {\special{em:lineto}} \put(31.92,24.52)
{\special{em:lineto}} \put(30.98,22.63)  {\special{em:lineto}}
\put(30.00,20.78)  {\special{em:lineto}} \put(29.00,18.96)
{\special{em:lineto}} \put(28.02,17.19)  {\special{em:lineto}}
\put(27.08,15.48)  {\special{em:lineto}} \put(26.21,13.83)
{\special{em:lineto}} \put(25.42,12.26)  {\special{em:lineto}}
\put(24.75,10.77)  {\special{em:lineto}} \put(24.21,9.380)
{\special{em:lineto}} \put(23.82,8.092)  {\special{em:lineto}}
\put(23.57,6.916)  {\special{em:lineto}} \put(23.57,4.916)
{\special{em:lineto}} }
\put(20,10){\unitlength=1mm\special{em:linewidth 0.3pt}
\put(23.57,4.916)  {\special{em:moveto}} \put(23.82,4.092)
{\special{em:lineto}} \put(24.21,3.378)  {\special{em:lineto}}
\put(24.75,2.769)  {\special{em:lineto}} \put(25.42,2.257)
{\special{em:lineto}} \put(26.21,1.830)  {\special{em:lineto}}
\put(27.08,1.479)  {\special{em:lineto}} \put(28.02,1.192)
{\special{em:lineto}} \put(29.00,0.961)  {\special{em:lineto}}
\put(30.00,0.776)  {\special{em:lineto}} \put(30.98,0.630)
{\special{em:lineto}} \put(31.92,0.514)  {\special{em:lineto}}
\put(32.77,0.425)  {\special{em:lineto}} \put(33.46,0.357)
{\special{em:lineto}} \put(33.90,0.308)  {\special{em:lineto}}
\put(33.98,0.276)  {\special{em:lineto}} \put(33.39,0.264)
{\special{em:lineto}} \put(31.49,0.290)  {\special{em:lineto}}
\put(30.80,0.305)  {\special{em:lineto}} \put(29.94,0.326)
{\special{em:lineto}} \put(28.82,0.358)  {\special{em:lineto}}
\put(27.31,0.410)  {\special{em:lineto}} \put(24.98,0.511)
{\special{em:lineto}} \put(24.00,0.6)  {\special{em:lineto}}
\put(22.00,0.7)  {\special{em:lineto}} \put(20.00,0.9)
{\special{em:lineto}} }
\put(20,10){\unitlength=1mm\special{em:linewidth 0.3pt}
\put(2.364,39.00) {\special{em:moveto}} \put(6.048,38.50)
{\special{em:lineto}} \put(8.552,38.00) {\special{em:lineto}}
\put(10.42,37.50) {\special{em:lineto}} \put(11.88,37.00)
{\special{em:lineto}} \put(13.06,36.50) {\special{em:lineto}}
\put(14.02,36.00) {\special{em:lineto}} \put(15.54,35.00)
{\special{em:lineto}} \put(16.66,34.00) {\special{em:lineto}}
\put(17.50,33.00) {\special{em:lineto}} \put(18.14,32.00)
{\special{em:lineto}} \put(18.63,31.00) {\special{em:lineto}}
\put(19.02,30.00) {\special{em:lineto}} \put(19.53,28.00)
{\special{em:lineto}} \put(19.81,26.00) {\special{em:lineto}}
\put(19.94,24.00) {\special{em:lineto}} \put(19.99,22.00)
{\special{em:lineto}} \put(20.00,20.00) {\special{em:lineto}}
\put(20.01,18.00) {\special{em:lineto}} \put(20.06,16.00)
{\special{em:lineto}} \put(20.19,14.00) {\special{em:lineto}}
\put(20.47,12.00) {\special{em:lineto}} \put(20.98,10.00)
{\special{em:lineto}} \put(21.37,9.00) {\special{em:lineto}}
\put(21.86,8.00) {\special{em:lineto}} \put(22.50,7.00)
{\special{em:lineto}} \put(23.34,6.00) {\special{em:lineto}}
\put(24.46,5.00) {\special{em:lineto}} \put(25.98,4.00)
{\special{em:lineto}} \put(28.12,3.00) {\special{em:lineto}}
\put(29.06,2.667) {\special{em:lineto}} \put(30.15,2.333)
{\special{em:lineto}} \put(31.45,2.00) {\special{em:lineto}}
\put(33.02,1.667) {\special{em:lineto}} \put(35.01,1.333)
{\special{em:lineto}} \put(37.63,1.000) {\special{em:lineto}}
\put(41.44,0.667) {\special{em:lineto}} } \end{picture}}

\vskip 9mm \unitlength=1mm
\centerline{\begin{picture}(82,57)\put(-2,59){\it b\hs{\rm )}}
\put(45.7,42.5){\it 1}\put(42.5,35.3){\it 2} \put(51,12.2){\it
1}\put(53.5,19.5){\it 2}
\put(-2,10){\vector(1,0){87}}\put(77,5){$\varepsilon-\mu$}
\put(20,10){\vector(0,1){51}}\put(22,59.5){$w$}\put(17,51.5){1}
\put(0,10){\line(0,-1){1}}\put(-2,5){$-\hh I$}
\put(20,10){\line(0,-1){1}}\put(19.3,5){$0$}
\multiput(60,11)(0,1.98){20}{\line(0,1){1.4}}
\put(60,10){\line(0,-1){1}}\put(55.9,5){$J-I$}
\multiput(-2,50)(2.02,0){31}{\line(1,0){1.4}}
\put(40,30){\circle*{0.7}}
\multiput(39.5,30)(-2,0){10}{\line(-1,0){1.4}}
\put(20,10){\line(-1,0){1}}\put(17,29){$\frac12$}
\multiput(40,29.5)(0,-2){10}{\line(0,-1){1.4}}
\put(40,10){\line(0,-1){1}}\put(32.3,5){$\frac12(J-I\hh )$}
\put(20,10){\unitlength=1mm\special{em:linewidth 0.3pt}
\put(20.00,4.500)  {\special{em:moveto}} \put(18.01,5.332)
{\special{em:lineto}} \put(16.73,6.044)  {\special{em:lineto}}
\put(15.90,6.616)  {\special{em:lineto}} \put(14.74,7.604)
{\special{em:lineto}} \put(13.95,8.500)  {\special{em:lineto}}
\put(13.64,8.924)  {\special{em:lineto}} \put(13.14,9.752)
{\special{em:lineto}} \put(12.60,10.94)  {\special{em:lineto}}
\put(12.10,12.82)  {\special{em:lineto}} \put(11.90,14.62)
{\special{em:lineto}} \put(11.89,16.34)  {\special{em:lineto}}
\put(12.00,18.01)  {\special{em:lineto}} \put(12.19,19.62)
{\special{em:lineto}} \put(12.42,21.16)  {\special{em:lineto}}
\put(12.66,22.64)  {\special{em:lineto}} \put(12.90,24.04)
{\special{em:lineto}} \put(13.10,25.38)  {\special{em:lineto}}
\put(13.28,26.65)  {\special{em:lineto}} \put(13.41,27.84)
{\special{em:lineto}} \put(13.50,28.96)  {\special{em:lineto}}
\put(13.50,30.96)  {\special{em:lineto}} }
\put(20,10){\unitlength=1mm\special{em:linewidth 0.3pt}
\put(13.50,30.96)  {\special{em:moveto}} \put(13.50,30.96)
{\special{em:lineto}} \put(13.41,31.84)  {\special{em:lineto}}
\put(13.28,32.65)  {\special{em:lineto}} \put(13.10,33.38)
{\special{em:lineto}} \put(12.90,34.04)  {\special{em:lineto}}
\put(12.66,34.64)  {\special{em:lineto}} \put(12.42,35.16)
{\special{em:lineto}} \put(12.19,35.62)  {\special{em:lineto}}
\put(12.00,36.01)  {\special{em:lineto}} \put(11.89,36.34)
{\special{em:lineto}} \put(11.90,36.62)  {\special{em:lineto}}
\put(12.10,36.82)  {\special{em:lineto}} \put(12.60,36.94)
{\special{em:lineto}} \put(13.64,36.92)  {\special{em:lineto}}
\put(13.95,36.90)  {\special{em:lineto}} \put(14.74,36.80)
{\special{em:lineto}} \put(15.26,36.73)  {\special{em:lineto}}
\put(15.90,36.62)  {\special{em:lineto}} \put(16.73,36.44)
{\special{em:lineto}} \put(18.01,36.13)  {\special{em:lineto}}
\put(20.00,35.50)  {\special{em:lineto}} }
\put(20,10){\unitlength=1mm\special{em:linewidth 0.3pt}
\put(20.00,35.50)  {\special{em:moveto}} \put(21.99,34.67)
{\special{em:lineto}} \put(23.27,33.96)  {\special{em:lineto}}
\put(24.10,33.38)  {\special{em:lineto}} \put(24.74,32.87)
{\special{em:lineto}} \put(25.26,32.40)  {\special{em:lineto}}
\put(25.69,31.94)  {\special{em:lineto}} \put(26.36,31.08)
{\special{em:lineto}} \put(27.40,29.06)  {\special{em:lineto}}
\put(27.90,27.18)  {\special{em:lineto}} \put(28.10,25.38)
{\special{em:lineto}} \put(28.11,23.66)  {\special{em:lineto}}
\put(28.00,21.99)  {\special{em:lineto}} \put(27.81,20.38)
{\special{em:lineto}} \put(27.58,18.84)  {\special{em:lineto}}
\put(27.34,17.36)  {\special{em:lineto}} \put(27.10,15.96)
{\special{em:lineto}} \put(26.90,14.62)  {\special{em:lineto}}
\put(26.72,13.35)  {\special{em:lineto}} \put(26.59,12.16)
{\special{em:lineto}} \put(26.50,11.04)  {\special{em:lineto}}
\put(26.50,9.040)  {\special{em:lineto}} }
\put(20,10){\unitlength=1mm\special{em:linewidth 0.3pt}
\put(26.50,9.040)  {\special{em:moveto}} \put(26.59,8.156)
{\special{em:lineto}} \put(26.72,7.348)  {\special{em:lineto}}
\put(26.90,6.616)  {\special{em:lineto}} \put(27.10,5.956)
{\special{em:lineto}} \put(27.34,5.364)  {\special{em:lineto}}
\put(27.58,4.840)  {\special{em:lineto}} \put(27.81,4.384)
{\special{em:lineto}} \put(28.00,3.988)  {\special{em:lineto}}
\put(28.11,3.655)  {\special{em:lineto}} \put(28.10,3.384)
{\special{em:lineto}} \put(27.90,3.183)  {\special{em:lineto}}
\put(27.40,3.065)  {\special{em:lineto}} \put(26.36,3.074)
{\special{em:lineto}} \put(24.10,3.384)  {\special{em:lineto}}
\put(23.27,3.554)  {\special{em:lineto}} \put(21.99,3.866)
{\special{em:lineto}} \put(20.00,4.500)  {\special{em:lineto}} }
\put(20,10){\unitlength=1mm\special{em:linewidth 0.3pt}
\put(-15.95,39.00) {\special{em:moveto}} \put(-10.18,38.50)
{\special{em:lineto}} \put(-6.166,38.00)  {\special{em:lineto}}
\put(-3.121,37.50)  {\special{em:lineto}} \put(-0.685,37.00)
{\special{em:lineto}} \put(1.332,36.50)  {\special{em:lineto}}
\put(3.042,36.00)  {\special{em:lineto}} \put(5.811,35.00)
{\special{em:lineto}} \put(7.981,34.00)  {\special{em:lineto}}
\put(9.744,33.00)  {\special{em:lineto}} \put(11.21,32.00)
{\special{em:lineto}} \put(12.45,31.00)  {\special{em:lineto}}
\put(13.52,30.00)  {\special{em:lineto}} \put(15.29,28.00)
{\special{em:lineto}} \put(16.71,26.00)  {\special{em:lineto}}
\put(17.92,24.00)  {\special{em:lineto}} \put(18.99,22.00)
{\special{em:lineto}} \put(20.00,20.00)  {\special{em:lineto}}
\put(21.01,18.00)  {\special{em:lineto}} \put(22.08,16.00)
{\special{em:lineto}} \put(23.29,14.00)  {\special{em:lineto}}
\put(24.71,12.00)  {\special{em:lineto}} \put(26.48,10.00)
{\special{em:lineto}} \put(27.55,9.00)  {\special{em:lineto}}
\put(28.79,8.00)  {\special{em:lineto}} \put(30.26,7.00)
{\special{em:lineto}} \put(32.02,6.00)  {\special{em:lineto}}
\put(34.19,5.00)  {\special{em:lineto}} \put(36.96,4.00)
{\special{em:lineto}} \put(40.69,3.00)  {\special{em:lineto}}
\put(46.17,2.00)  {\special{em:lineto}} \put(55.95,1.00)
{\special{em:lineto}} \put(61.83,0.667)  {\special{em:lineto}} }
\put(20,10){\unitlength=1mm\special{em:linewidth 0.3pt}
\put(20.00,12.50)  {\special{em:moveto}} \put(19.23,13.13)
{\special{em:lineto}} \put(18.11,14.61)  {\special{em:lineto}}
\put(17.53,15.67)  {\special{em:lineto}} \put(17.12,16.60)
{\special{em:lineto}} \put(16.82,17.45)  {\special{em:lineto}}
\put(16.58,18.25)  {\special{em:lineto}} \put(16.40,19.00)
{\special{em:lineto}} \put(16.26,19.72)  {\special{em:lineto}}
\put(16.14,20.40)  {\special{em:lineto}} \put(16.05,21.06)
{\special{em:lineto}} \put(15.98,21.68)  {\special{em:lineto}}
\put(15.92,22.29)  {\special{em:lineto}} \put(15.88,22.87)
{\special{em:lineto}} \put(15.85,23.43)  {\special{em:lineto}}
\put(15.83,23.96)  {\special{em:lineto}} \put(15.83,24.96)
{\special{em:lineto}} }
\put(20,10){\unitlength=1mm\special{em:linewidth 0.3pt}
\put(15.83,24.96)  {\special{em:moveto}} \put(15.85,25.43)
{\special{em:lineto}} \put(15.88,25.87)  {\special{em:lineto}}
\put(15.92,26.29)  {\special{em:lineto}} \put(15.98,26.68)
{\special{em:lineto}} \put(16.05,27.06)  {\special{em:lineto}}
\put(16.14,27.40)  {\special{em:lineto}} \put(16.26,27.72)
{\special{em:lineto}} \put(16.40,28.00)  {\special{em:lineto}}
\put(16.58,28.25)  {\special{em:lineto}} \put(16.82,28.45)
{\special{em:lineto}} \put(17.12,28.60)  {\special{em:lineto}}
\put(17.53,28.67)  {\special{em:lineto}} \put(18.11,28.61)
{\special{em:lineto}} \put(19.23,28.13)  {\special{em:lineto}}
\put(20.00,27.50)  {\special{em:lineto}} }
\put(20,10){\unitlength=1mm\special{em:linewidth 0.3pt}
\put(20.00,27.50)  {\special{em:moveto}} \put(20.77,26.87)
{\special{em:lineto}} \put(21.89,25.39)  {\special{em:lineto}}
\put(22.47,24.33)  {\special{em:lineto}} \put(22.88,23.40)
{\special{em:lineto}} \put(23.18,22.55)  {\special{em:lineto}}
\put(23.42,21.75)  {\special{em:lineto}} \put(23.60,21.00)
{\special{em:lineto}} \put(23.74,20.28)  {\special{em:lineto}}
\put(23.86,19.60)  {\special{em:lineto}} \put(23.95,18.94)
{\special{em:lineto}} \put(24.02,18.32)  {\special{em:lineto}}
\put(24.08,17.71)  {\special{em:lineto}} \put(24.12,17.13)
{\special{em:lineto}} \put(24.15,16.57)  {\special{em:lineto}}
\put(24.17,16.04)  {\special{em:lineto}} \put(24.17,15.04)
{\special{em:lineto}} }
\put(20,10){\unitlength=1mm\special{em:linewidth 0.3pt}
\put(24.17,15.04)  {\special{em:moveto}} \put(24.15,14.57)
{\special{em:lineto}} \put(24.12,14.13)  {\special{em:lineto}}
\put(24.08,13.71)  {\special{em:lineto}} \put(24.02,13.32)
{\special{em:lineto}} \put(23.95,12.94)  {\special{em:lineto}}
\put(23.86,12.60)  {\special{em:lineto}} \put(23.74,12.28)
{\special{em:lineto}} \put(23.60,12.00)  {\special{em:lineto}}
\put(23.42,11.75)  {\special{em:lineto}} \put(23.18,11.55)
{\special{em:lineto}} \put(22.88,11.40)  {\special{em:lineto}}
\put(22.47,11.33)  {\special{em:lineto}} \put(21.89,11.39)
{\special{em:lineto}} \put(20.77,11.87)  {\special{em:lineto}}
\put(20.00,12.50)  {\special{em:lineto}} }
\put(20,10){\unitlength=1mm\special{em:linewidth 0.3pt}
\put(-22.00,38.20) {\special{em:moveto}} \put(-17.94,38.00)
{\special{em:lineto}} \put(-13.95,37.50) {\special{em:lineto}}
\put(-10.74,37.00) {\special{em:lineto}} \put(-5.747,36.00)
{\special{em:lineto}} \put(-1.972,35.00) {\special{em:lineto}}
\put(1.042,34.00) {\special{em:lineto}} \put(5.661,32.00)
{\special{em:lineto}} \put(9.128,30.00) {\special{em:lineto}}
\put(11.90,28.00) {\special{em:lineto}} \put(14.24,26.00)
{\special{em:lineto}} \put(16.30,24.00) {\special{em:lineto}}
\put(18.18,22.00) {\special{em:lineto}} \put(20.00,20.00)
{\special{em:lineto}} \put(21.82,18.00) {\special{em:lineto}}
\put(23.70,16.00) {\special{em:lineto}} \put(25.76,14.00)
{\special{em:lineto}} \put(28.10,12.00) {\special{em:lineto}}
\put(30.87,10.00) {\special{em:lineto}} \put(32.50,9.000)
{\special{em:lineto}} \put(34.34,8.000) {\special{em:lineto}}
\put(36.46,7.000) {\special{em:lineto}} \put(38.96,6.000)
{\special{em:lineto}} \put(41.98,5.000) {\special{em:lineto}}
\put(43.13,4.667) {\special{em:lineto}} \put(44.38,4.333)
{\special{em:lineto}} \put(45.74,4.000) {\special{em:lineto}}
\put(47.24,3.667) {\special{em:lineto}} \put(48.90,3.333)
{\special{em:lineto}} \put(50.74,3.000) {\special{em:lineto}}
\put(52.81,2.667) {\special{em:lineto}} \put(55.18,2.333)
{\special{em:lineto}} \put(57.94,2.000) {\special{em:lineto}}
\put(61.24,1.667) {\special{em:lineto}} } \end{picture}}

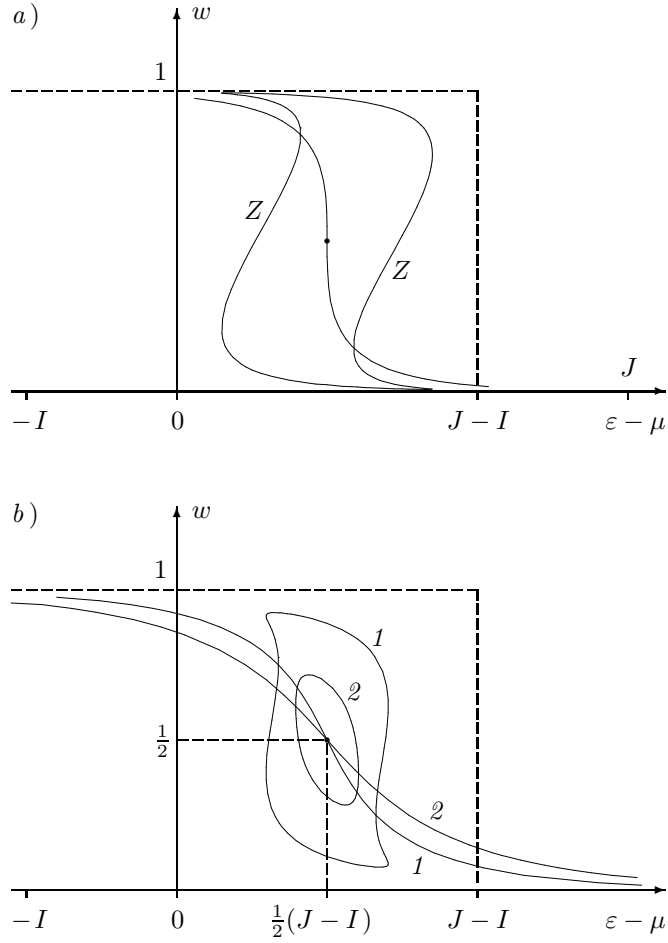
\centerline{\it Fig. 3. Probability $w$ as a function of $x=(\varepsilon-\mu)/I$ for different values of the parameter $\theta=T/T^{\h\pr}_C$: }
\centerline{\it a{\rm )} $\theta=0.5$;
{\it b\hh{\rm )}} 1 -- $\theta=0.75$; 2 -- $\theta=0.95$. }\vskip 3mm

\par The order parameter equation (53) can be easily obtained from equations (37):
$$ \ln\hh\frac{1+\xi}{1-\xi}=\frac{2}{\theta}\hs\xi\hh , $$
from which the approximate equations can be derived
$$ \left.\begin{array}{ccc}
\xi\approx 1-2\hs e^{\h -\h 2/\theta} & \hbox{at} & \theta\ll 1
\hh , \medskip \\
\xi\approx \sqrt{\h 3\hh(1-\theta)} & \hbox{at} & \theta\leq 1
\hh . \\ \end{array}\right. $$

\par The distribution function (59) discontinuity at $E=E_2$ determines the existence of a gap in the electron energy spectrum. For the gap parameter
$$ \triangle=\frac{I}{2}\hh(w_2-w_1)|_{E\h =\h E_2} $$
the following approximate equations can be obtained:
$$ \left.\begin{array}{ccc}\ds\frac{\triangle}{\triangle_{\h\rm o}}=
1-\frac{4\hh\triangle_{\h\rm o}}{k_B\h T}\hs
e^{\h -\h 2\h\triangle_{\h\rm o}/k_B\h T} & \hbox{at} & T\ll T_C
\hh , \medskip \\
\ds\frac{\triangle}{\triangle_{\rm o}}=\sqrt{1-\ds\frac{T}{T_C}} & \hbox{at} & T\leq T_C \hh , \\ \end{array}\right. $$
where $\triangle_{\h\rm o}=\triangle(T=0)=1,72\hs k_B\hh T_C$.
These equations are in agreement with experimental temperature dependencies of the gap parameter.

\par The state of the electronic system described by an anisotropic distribution function at $T_C\leq T<T^\pr_C$ is unstable. And only at $T<T_C$, the state becomes stable together with the appearance of the gap in the energy spectrum.

\vskip 4mm
\centerline{\bf Conclusions }\vskip 2mm

\par The modern microscopic theory of superconductivity developed by Bardeen, Cooper and Schrieffer [8] is based on the assumption that the superconductivity is caused by the effective attraction between electrons with the wave vectors lying near the Fermi surface. This attraction leads to creation pairs of electrons with opposite momenta and spins. In this case, the Coulomb repulsion between the electrons prevents them from pairing and thereby prohibit the electronic system transition to the superconducting state. Therefore, the prevalence of the effective electron attraction over the repulsion at least for some values of the wave vectors is necessary for the existence of superconductivity. In the theory of superconductivity, there is considered a model of electrons with quadratic dispersion law, the effective attraction of which is described by means of some simplified Hamiltonian. The mathematical formulation of the theory is carried out using the methods of quantum field theory.

\par Frohlich and Bardeen showed that the effective attraction between electrons can occur when the electrons interact with crystal lattice vibrations. In the framework of the quantum theory, such interaction is represented as the emission and absorption of phonons by electrons.

\par In connection with the discovery of high-temperature superconductivity, there was appeared the necessity to explain the reasons for strong electron pairing which causes high critical temperature. For this purpose, various nonphonon mechanisms of the electron attraction efficiency have been proposed. The proposed mechanisms differ from each other mainly in the type of particles by means of which the electrons interact, and the nature of the pairing. Apart from phonons, excitons, plasmons, magnons and other particles were considered as particles realizing the effective attraction between electrons.

\par The model of electrons in metals used in the present work can be the basis for an alternative theory of superconductivity. This model significantly differs from the models used in the modern theory of superconductivity. Here, the superconductivity is caused by the repulsion between electrons with the wave vectors $\bf k$ and $-\hh\bf k^\pr$, and the energy gap in the spectrum is created due to the attraction between electrons with identical wave vectors, which is a consequence of the exchange interaction. The statistical description of electrons is carried out within the framework of the density matrix formalism, the distinguishing feature of which is its simplicity typical for both the mathematical equations form and their physical content. The considered problem demonstrates the advantages of the density matrix method.

\par The results of this work were obtained under the assumption that the interaction energy of electrons with the wave vectors ${\bf k}$ and
${\bf k}^\pr$ can be approximated by the expression
$$ V_{{\bf kk}^\pr}=-\hh I\hh\delta_{{\bf k}\h -\h{\bf k}^\pr}+
J\hh\delta_{{\bf k}\h +\h{\bf k}^\pr}\hh . $$

\par The use of this approximation equation allows to find the solutions of equation (28) in a sufficiently simple form and prove the principal possibility of existence of the aforesaid distribution function features. Of course, some questions remain unresolved: how strongly the approximate expression differs from the real energy of the electron interaction and whether the considered distribution function features are only a consequence of the approximation used here. A complete answer to the last question can be found only after a comprehensive study of different solutions of equations (28), in which a more accurate expression of the type (34) or (35) is substituted instead of the approximate expression (36) for the one-electron energy. The solution of the nonlinear integral equation obtained in this way represents a very difficult problem that can be solved only by numerical methods. However, as noted above, the used approximation does not distort the character of the electron interaction energy dependence on wave vectors. It gives reasons to assume that the described features of the distribution function are preserved if expression (36) in equation (28) is replaced by the more precise one.

\par The temperature of the electron gas transition to the superconducting state is determined by the average energy values $I$ and $J$ of the electron interaction. Equations (32) and (33) for the electron interaction energies are very rough and cannot be used for accurate calculations of the critical temperature values. Because these equations do not take into account oscillations of the crystal lattice. They are obtained under the assumption that the conduction electron motion is described by a Bloch wave function in the entire volume of the crystal. The characteristic spatial size of the wave function is determined by the coherence length $\xi_{\rm o}$; its typical value is equal to
$10^{\hh 3}$ $\stackrel\circ A$. As is known, the exact calculation of the electron-electron interaction energy is very complicated. However, using equation (32), one can estimate the interaction energy value for two electrons with the same wave vectors: $I\sim {e^2}/\xi_{\rm o}=10^{\h -\h 22}$J. The critical temperature $T_C=I/2\hh k_B\hh \sim 5$ K corresponds to this energy value.  This temperature coincides by order of magnitude with the transition temperatures of the chemical elements that show superconductivity.

\vskip 4mm

\par Summarizing, let us enumerate the main results of the work:

\par 1. A new method is proposed for the statistical description of the equilibrium many-particle system by means of a density matrix, which is a generalization of the Hartree - Fock - Slater variational method.

\par 2. This method is used for description of conduction electrons in metals. An integral equation for the function of electron distribution over wave vectors has been obtained. It is shown that the distribution function has previously unknown features which are expressed in the fact that for the electrons in the lowest energy state, the function is multi-valued, anisotropic ($w_{\h -\h{\bf k}}\neq w_{\bf k}$) and has a discontinuity on one of the isoenergetic surfaces, in a thin layer adjacent to the Fermi surface.

\par 3. It was suggested that superconductivity can be interpreted as a consequence of anisotropy in the electron distribution over wave vectors and the presence of a gap in the energy spectrum of the conduction electrons is connected with the distribution function discontinuity [9].

\vskip 4mm

\def\qhh{\hskip -5.7mm\qquad\parbox[t][1\height]{173mm}}

\centerline{\bf References }\vskip 2mm

\noindent [1] \qhh{ J. von Neumann, Mathematical Foundations of Quantum Mechanics, Moscow: Nauka, 1964. } \vskip 1.5mm

\noindent [2] \qhh{ K. Blum, Density Matrix Theory and Applications, Moscow: Mir, 1983. } \vskip 1.5mm

\noindent [3] \qhh{ N.N. Bogolyubov, N.N. Bogolyubov, Introduction to quantum statistical mechanics, Moscow: Nauka, 1984. } \vskip 1.5mm

\noindent [4] \qhh{ B.V. Bondarev, Density matrix method in quantum theory of cooperative process, Moscow: Publishing company Sputnik$^+$, 2001, p. 250. } \vskip 1.5mm

\noindent [5] \qhh{ B.V. Bondarev, {\it Quantum markovian master equation for system of identical particles interacting with a heat reservoir, } Рhysisa А, 1991, v. 176, р. 366-386. } \vskip 1.5mm

\noindent [6] \qhh{ B.V. Bondarev, {\it Quantum lattice gas. Method of
density matrix, } Рhysisa А, 1992, v. 184, р. 205-230. } \vskip 1.5mm

\noindent [7] \qhh{ N. Ashcroft, N. Mermin, Solid State Physics, Moscow: Mir, 1979. } \vskip 1.5mm

\noindent [8] \qhh{ J. Bardeen, L. Cooper, J. Schrieffer, Theory of superconductivity, under the editorship of N.N.Bogolyubov, Moscow, IL, 1960, p. 103-171. } \vskip 1.5mm

\noindent [9] \qhh{ B.V. Bondarev, {\it On some peculiarities of electrons distribution function over the Bloch states, } Vestnik MAI, 1996, vol. 3, № 2, p. 56 - 65. }\vskip 1.5mm

\end{document}